\pdfoutput=1

\documentclass[a4paper,11pt]{article}

\usepackage{jheppub,fancyhdr,epstopdf,color,amsmath,cases,slashed,hyperref,subfigure}
\usepackage{amsfonts}
\usepackage{amssymb}
\usepackage{graphicx,graphics,epsfig}
\usepackage{geometry}
\usepackage[english]{babel}
\usepackage{mathtools}
\usepackage[utf8]{inputenc}
\usepackage[T1]{fontenc}
 \usepackage{float}
\usepackage{hyperref}
\usepackage{caption}
\usepackage{caption}
\usepackage{subfigure}  
\usepackage{mathrsfs}  
\usepackage{comment}
\usepackage{float}
\usepackage{stackrel}
\usepackage{multirow}
\usepackage{slashed}

\def\noi{\noindent}
\def\non{\nonumber}

\def\l{\lambda}

\newcommand{\Lag}{\mathcal{L}}

\def\ba{\begin{array}}
\def\ea{\end{array}}
\def\bea{\begin{eqnarray}}
\def\eea{\end{eqnarray}}
\newcommand{\Omegah}{\Omega h^{2}}

\newcommand{\Asla}{\not{\hbox{\kern-3.5pt $A$}}}
\newcommand{\Gsla}{\not{\hbox{\kern-3.5pt $G$}}}
\newcommand{\Wsla}{\not{\hbox{\kern-3.5pt $W$}}}
\newcommand{\Zsla}{\not{\hbox{\kern-3.5pt $Z$}}}

\newcommand{\Dslash}{\not{\hbox{\kern-4pt $D$}}}
\newcommand{\pslash}{\not{\hbox{\kern-2.3pt $p$}}}

\def\lsim{\;\raise0.3ex\hbox{$<$\kern-0.75em\raise-1.1ex\hbox{$\sim$}}\;}
\def\gsim{\;\raise0.3ex\hbox{$>$\kern-0.75em\raise-1.1ex\hbox{$\sim$}}\;}

\def\l{\lambda}

\def\ba{\begin{array}}
\def\ea{\end{array}}
\def\bea{\begin{eqnarray}}
\def\eea{\end{eqnarray}}

\def\bll{\tilde{\beta}_{\l_L}}

\def\lsim{\;\raise0.3ex\hbox{$<$\kern-0.75em\raise-1.1ex\hbox{$\sim$}}\;}
\def\gsim{\;\raise0.3ex\hbox{$>$\kern-0.75em\raise-1.1ex\hbox{$\sim$}}\;}

\newcommand{\beqn}{\begin{eqnarray}}
\newcommand{\eeqn}{\end{eqnarray}}

\title{Relic density of dark matter in the inert doublet model beyond leading order for the low mass region: 1. Renormalisation and constraints }
\preprint{LAPTH-001/21, CERN-TH-2021-001}

\author[a]{Shankha Banerjee}
\author[b]{\!\!, Fawzi Boudjema}
\author[c, d]{\!\!, Nabarun Chakrabarty}
\author[e]{\!\!,  Hao Sun}

\affiliation[a]{CERN, Theoretical Physics Department, CH-1211 Geneva 23, Switzerland}
\affiliation[b]{LAPTh, Universit\'e Savoie Mont Blanc, CNRS, BP~110, F-74941 Annecy-le-Vieux, France}
\affiliation[c]{Centre for High Energy Physics, Indian Institute of Science, C.V. Raman Avenue, Bangalore 560012, India}
\affiliation[d]{Department of Physics, Indian Institute of Technology Kanpur, Kanpur, Uttar Pradesh 208016, India}
\affiliation[e]{Institute of Theoretical Physics, School of Physics, Dalian University of Technology, Dalian 116024, People’s Republic of China}

\emailAdd{shankha.banerjee@cern.ch}
\emailAdd{boudjema@lapth.cnrs.fr}
\emailAdd{chakrabartynabarun@gmail.com}
\emailAdd{haosun@dlut.edu.cn}

\abstract{The present paper is the first in a series that addresses the calculation of  the full one-loop corrections of dark matter (DM) annihilation cross-sections in the low mass region of the inert doublet model (IDM). This series is a sequel to our recent publication concerning these corrections in the high mass region. We first review the renormalisation of the model both in a fully on-shell (OS) scheme and a mixed scheme combining on-shell (for the masses) and a $\overline{{\rm MS}}$ approach when the partial invisible width is closed and does not allow the use of a full OS scheme. The scale dependence introduced by the mixed scheme is shown to be tracked through an analysis of  a parametrisation of the tree-level cross-section and the $\beta$ constant of a specific coupling. We discuss how to minimise the scale dependence. The theoretical uncertainty brought by the scale dependence leads us to introduce a new criterion on the perturbativity of the IDM. This criterion further delimits  the allowed parameter space which we investigate carefully by including a host of constraints, both theoretical and experimental, including in particular, new data from the LHC. We come up with a set of benchmark points that cover three different mechanisms for a viable relic density of DM: {\it i)} a dominance of co-annihilation into a fermion pair, { \it ii)} annihilation into 2 vector bosons of which one is off-shell that requires the calculation of a $2 \to 3$ process, { \it iii)} annihilation that proceeds through the very narrow standard model Higgs resonance. Since the $2 \to 3$ vector boson channel features in all three channels and is essentially a build up on the simpler annihilation to OS vector bosons, we study the latter in detail in the present paper. We confirm again that the corrected cross-sections involve a parameter that represents rescattering in the dark sector that a tree-level computation in not sensitive to.
The set-up of the renormalisation detailed in the present paper will be the backbone of the accompanying papers where each mechanism requires a specific treatment. One-loop corrections to a $2 \to 3$ process for DM annihilation is technically challenging and has not been attempted before. We dedicate one of the accompanying papers to such a computation. The one-loop correction in the presence of a resonance will be presented separately since we need to  supplement our general schemes with a complex scheme.}

\begin{document}

\date\today

\maketitle


\section{Introduction}
\label{sec:intro}
Cosmology has entered the era of precision measurements. One such measurement is the inferred relic density of dark matter (DM) that is now determined at the per-cent level~\cite{Ade:2015xua}. For a particle physicist such an accuracy is reminiscent of the one achieved at LEP. In turn, on the theory side,  this level of accuracy requires that observables be computed with a precision on par with the experimental precision or even better. Given the cosmological model for the thermodynamics/evolution of the universe  and a model of DM, the computation of the relic density of DM involves the calculations of annihilation rates. While, unfortunately, no  sign of a model of DM has emerged either in direct detection or at the colliders, many models have been proposed and a huge amount of work has been dedicated to the search and study of these models. Yet, despite the central role that the very precise measurement of the relic density plays in constraining the phenomenology of these models, only very few examples have provided the calculation of the annihilation rates beyond the leading order, tree-level, approximation. In these series of papers, we will tackle the computation of many annihilation rates that occur in different viable scenarios of the  the inert double model (IDM)~\cite{Deshpande:1977rw, Barbieri:2006dq,Hambye:2007vf}, in particular the low mass DM region in this model, beyond the tree-level approximation. This is a continuation of the work we initiated in~\cite{Banerjee:2019luv} to cover the heavy DM scenario, $500$ GeV to $1$ TeV, which complements important non-perturbative electroweak Sommerfeld effects~\cite{Hisano:2002fk, Hisano:2003ec, Hisano:2004ds, Hambye:2009pw, Biondini:2017ufr} that are relevant beyond the TeV.\\

\noindent The IDM, with its possible link between the Higgs sector and DM~\cite{Hambye:2007vf} has enjoyed some popularity~\cite{LopezHonorez:2006gr, Cao:2007rm, Agrawal:2008xz, Hambye:2009pw, Lundstrom:2008ai, Andreas:2009hj, Arina:2009um, Dolle:2009ft, Nezri:2009jd, Miao:2010rg, Gong:2012ri, Gustafsson:2012aj, Swiezewska:2012eh, Wang:2012zv, Goudelis:2013uca, Arhrib:2013ela, Krawczyk:2013jta, Osland:2013sla, Abe:2015rja, Arhrib:2015hoa, Blinov:2015qva, Diaz:2015pyv, Ilnicka:2015jba, Belanger:2015kga, Carmona:2015haa, Kanemura:2016sos,Queiroz:2015utg,Belyaev:2016lok,Arcadi:2019lka, Eiteneuer:2017hoh, Ilnicka:2018def, Kalinowski:2018ylg,Basu:2020qoe,Abouabid:2020eik} but its structure beyond the tree-level has only been looked at either for some specific applications or conditions~\cite{Hambye:2007vf,Ferreira:2009jb,Ferreira:2015pfi,Kanemura:2002vm, Senaha:2018xek, Braathen:2019pxr, Arhrib:2012ia,Arhrib:2015hoa,Gustafsson:2007pc,Garcia-Cely:2015khw,Banerjee:2016vrp}. In order to perform one-loop calculations, for any process in the IDM and in particular for DM annihilation into standard model (SM) particles, we need a full and coherent renormalisation program for the IDM. We present the details of our renormalisation schemes and show many important features that we believe will be very useful not only in studying DM annihilation in the IDM but also in studying the renormalisation of models beyond the standard model (BSM). \\

The plan of this first paper in the series is as follows. We, in the next section, section~\ref{sec:tree-level}, briefly describe the model bringing forth the physical parameters (rather than the  parameters of the underlying Lagrangian). This will set the stage to the section, section~\ref{sec:renormalisation}, on renormalisation where we strive as much as possible to take an on-shell (OS) scheme, which will use the physical masses of the model and the partial width of the SM Higgs to a pair of DM. Even in the narrow range of masses for the IDM that we will study here, a fully OS scheme is not possible when the latter's partial width is kinematically closed. In this case we advocate a mixed OS-$\overline{MS}$ scheme. This mixed scheme introduces a scale dependence but we will show, on a simple example first, how to track the scale dependence through the $\beta$ constant of the associated coupling and a knowledge of the tree-level dependence on this coupling. This section will give us the opportunity to present our automated code to conduct one-loop calculations and the means we have to check the correctness of the results (switching between different choices of non-linear gauge-fixing parameters~\cite{Boudjema:1995cb,Baro:2007em} and  ultra-violet finiteness of the cross-sections and decays). Section~\ref{sec:constraints} is lengthy but necessary since we reanalyse all available experimental data and theoretical arguments to delimit the new  parameter space of the model. The constraints include new data on direct detection and also the LHC. We then propose a set of benchmark points which will then be scrutinised further by computing their respective relic density based on one-loop cross-sections. We will then see how the relic density predictions compare with those derived with tree-level cross-sections. This, thorough scan reveals in fact three mechanisms that permit a good DM candidate with a mass below the $W$-boson's, $M_W$. The features of these mechanisms are summarised in section~\ref{sec:summary_channels}. They consist of {\it i)} a narrow co-annihilation region driven essentially by gauge coupling, {\it ii)} annihilations driven essentially by the SM Higgs resonance and {\it iii)} annihilations into 3-body final states, $W f \bar f^\prime, Z f \bar f$ built up (mainly) on $W W^\star, Z Z^\star$ ($W^\star, Z^\star$ denote the off-shell vector bosons, a notion which will become clearer when we study the $2 \to 3$ processes). The technicalities involved in these three mechanisms are quite different and this is the reason we decided to present the calculations in these three cases in three separate publications for more clarity and readability. Both {\it ii)} and {\it iii)} are very challenging. This is the first time that annihilation of DM to 3 particles is conducted at one-loop. Renormalisation and loop corrections in the presence of a resonance require extreme care, in fact our paper~\cite{OurPaper4_2020} on the mechanism of annihilation through a Higgs can serve as a good example for many other models and processes. Very subtle issues about renormalisation beyond what is presented in this first paper of the series will be highlighted in~\cite{OurPaper4_2020}. Moreover, even in the so called co-annihilation scenario~\cite{OurPaper2_2020}, there is a small (but non-negligible) contribution that proceeds through annihilation to 3-body final state. A contribution from $W f \bar f^\prime$ will also feature in the Higgs resonance mechanism. This is one of the reasons why before tackling the challenging $W f \bar f^\prime, Z f \bar f$ in~\cite{OurPaper3_2020}, the present paper studies in section~\ref{sec:XXVVwarmup}, as a warm-up, the annihilation to a pair of on-shell $W^+ W^-$ and $ZZ$, even if phenomenologically they lead to under-abundance. Another reason is that in this mass range a fully OS scheme is not possible and we need to use the mixed scheme. We therefore study the scale dependence of these annihilation cross-sections and suggest an optimal scale for such processes thereby putting a conjecture we make for the Higgs decay in this paper and for other processes elsewhere, on more solid ground. This study will also reveal that certain choices of parameters lead to too large scale dependence and even a breakdown of perturbativity in the sense of the loop expansion. We propose a criterion to avoid such configurations, criterion that goes beyond the perturbativity argument used in our constraints in section~\ref{sec:constraints}. This helps reduce the number of the benchmark points we use in the accompanying papers of this series. We end the present paper with a short conclusion in section~\ref{sec:conclusions}.


\section{The model and the parameters}
\label{sec:tree-level}
The IDM consists, in addition to the Standard Model (SM) Higgs doublet $\Phi_1$, of an extra doublet of scalars $\Phi_2$ on which a discrete $\mathbb{Z}_2$ symmetry is imposed. This symmetry  entails that $\Phi_2$ is odd while all other fields (of the SM) are even. As a consequence, this symmetry guarantees the stability of the lightest of the scalars of the $\Phi_2$ doublet. If the latter is neutral it qualifies as a possible dark matter candidate. Another important upshot of this symmetry is that these extra scalar fields in $\Phi_2$ cannot couple to fermions, at least through renormalisable operators.  Keeping only renormalisable operators, the scalar sector is therefore modified  to 

\begin{equation}
\Lag_{IDM}^{\text{scalar}} = (D^\mu \Phi_1)^\dagger D_\mu \Phi_1 + (D^\mu \Phi_2)^\dagger D_\mu \Phi_2 -
\mathscr{V}_{IDM}(\Phi_1,\Phi_2), 
\end{equation}\noi

with

\begin{eqnarray}
\label{eq:IDMpot}
 \mathscr{V}_{IDM}(\Phi_1,\Phi_2) &=& \mu_1^2 |\Phi_1|^2 + \mu_2^2 |\Phi_2|^2 +
\l_1 |\Phi_1|^4 + \l_2 |\Phi_2|^4  \non \\
& & + \l_3 |\Phi_1|^2 |\Phi_2|^2 +
\l_4 (\Phi_2^\dagger \Phi_1)(\Phi_1^\dagger \Phi_2) +
\left(\frac{\l_5}{2}(\Phi_1^\dagger \Phi_2)^2 + \mbox{h.c} \right).
\end{eqnarray}\noi
$\mu_i$ and  $\l_i$ are real and $D_\mu$ is the covariant derivative.

We parameterise the doublets as 

\begin{equation}
 \Phi_1 = \begin{pmatrix}
           G^+ \\
	\frac{1}{\sqrt{2}}\left(v + h + i G^0\right)
          \end{pmatrix}~\mbox{and}~
\Phi_2 = \begin{pmatrix}
          H^+ \\
    \frac{1}{\sqrt{2}}\left(X + i A \right)
         \end{pmatrix},
\end{equation}\noi

where $v$ is the SM vacuum expectation value (vev) with $v \simeq 246$ GeV, defined from the measurement of the $W$ ($M_W$) and $Z$ ($M_Z$) masses. We have

\begin{equation}
\label{eq:def_sw2}
s_W^2 \equiv \sin^2 \theta_W = 1 - \frac{M_W^2}{M_Z^2}, \quad 
M_W = \frac{1}{2}\frac{e}{s_W} v, \quad \Bigg(v= \frac{2 M_W s_W}{e}\Bigg),
\end{equation}\noi

\noi where $e$ is the electromagnetic coupling. The  $SU(2)$ gauge coupling, $g$, and the hypercharge gauge coupling, $g^\prime$, are then

\beqn
g=e/s_W, \quad g^\prime=e/c_W.
\eeqn
$h$ is the SM  Higgs boson (with mass $M_h=125$ GeV), and $G^0,G^\pm$ are, respectively, the neutral and  the charged Goldstone bosons. $X$ and $A$ are the new neutral physical scalars~\footnote{Since these additional scalars do not couple to the fermions (of the SM), we can not assign them definite CP numbers. By an abuse of language, we will, nonetheless, call $A$ the pseudo-scalar.} and $H^\pm$ is the charged physical scalar. While both $X$ and $A$ are possible DM candidates, the physics is the same through the interchange $(\l_5, X)\leftrightarrow (-\l_5, A)$. In these series of papers we take, for definiteness, $X$ as the DM candidate. Note that, in order to avoid any confusion between the neutral scalars of the model, we have labelled the fields differently than in our previous paper~\cite{Banerjee:2019luv}.

It is much instructive to revert to the description of the model through the physical parameters, especially when we will be seeking, as much as possible, an on-shell renormalisation of the model based on physical observables. In particular, the parameters of the potential can be translated to the physical masses of the scalars  
\begin{align}
\label{hmass}
M_h^2 &=   2 \l_1 v^2, \\
\label{Hpmass}
M_{H^\pm}^2 &= \mu_2^2 + \l_3 \frac{v^2}{2},\\
 \label{Hmass}
M_X^2 &= \mu_2^2 + \l_L \frac{v^2}{2}=  M_{H^\pm}^2 +
\left(\l_4+\l_5 \right)  \frac{v^2}{2},\\
 \label{Amass}
M_A^2 &= \mu_2^2 + \l_A \frac{v^2}{2}= M_{H^\pm}^2 +
\left(\l_4-\l_5\right) \frac{v^2}{2}= M_X^2 - \l_5 v^2,\quad \text{where} \nonumber \\
\l_{L/A} &= \l_3+\l_4\pm \l_5.
\end{align}\noi 

\newcommand{\hxx}{{\cal{A}}_{hXX}^0}
\newcommand{\hxxt}{${\cal{A}}_{hXX}^0$}

Unfortunately, the masses do not provide enough input to determine all the independent parameters of the potential. First of all, $\l_2$ is a parameter that describes the interaction solely within the dark sector $X,A,H^\pm$. Therefore, \underline{at tree-level}, the SM particles are insensitive to this parameter even in their interaction with the dark sector particles $X,A,H^\pm$ particles. Nonetheless, as we will see, one-loop observables will depend on this parameter that describes scattering/re-scattering in the DM sector. Second, to fully define  an observable, we still need an extra parameter, $\mu_2$, $\l_3$, or alternatively $\l_L$ The latter seems to be the most appropriate since it has a direct physical interpretation. Indeed the coupling of the SM Higgs, $h$, to a pair of DM is (at the amplitude level) given by 
\beqn
\label{eq:deflambdaL}
\hxx=-\l_L\; v.
\eeqn
\noi The superscript $^{0}$ relates to the tree-level definition. {\it In lieu}  of the mass term, $\mu_2^2$, we take $\l_L$ as an input parameter. The parameters of the potential in equation~\ref{eq:IDMpot} can then be reconstructed from the {\it input} parameters ($M_h,M_X,M_A,M_H^\pm,\l_L$) as 
 
\beqn
\label{eq:livsms}
\l_1 &= &  \frac{M_h^2 }{2  v^2}, \quad \quad (\l_1 \sim + 0.129), \nonumber \\
\l_5&=&\frac{M_X^2-M_A^2}{v^2}, \nonumber \\
\l_4&=& \l_5+ 2 \frac{M_A^2-M_{H^\pm}^2}{v^2},\nonumber \\
\l_3&=&\l_L-\l_4-\l_5 \\
\label{Xmass}
\Bigg( \mu_2^2&=&M_X^2 - \l_L \frac{v^2}{2}  \quad \to \quad \l_L=\frac{2(M_X^2 -\mu_2^2)}{v^2} \Bigg).
\eeqn
\noi $\mu_2$ is a redundant parameter. As mentioned earlier,  $\l_2$ is a parameter, which, at tree-level is not accessible since it describes the interactions solely within the dark sector. To define any benchmark and in view of the OS scheme for the renormalisation of the IDM that we advocate, apart from the SM parameters the input (physical) parameters we choose are 
\beqn
M_X, M_A, M_{H^\pm}, \l_L, (\l_2) 
\eeqn

It is instructive to make the following comments which will prove helpful later. When the DM candidate, $X$, is almost degenerate in mass with $A$, co-annihilation may be important. This scenario requires $\l_5 \sim 0$. In another limit, $M_{H^\pm}=M_A$, the electroweak custodial symmetry parameter, $T=0$ at one-loop even when $M_A,M_{H^\pm}$ are large. In this limit with $M_{A,H^\pm} > M_X$, $\l_4=\l_5=-\lambda<0$ (with $\lambda >0$). As we will see later, direct detection will impose that $\l_L \ll 1$ which means that $\l_3 \sim 2 \lambda$ when the custodial symmetry is imposed.


\section{Renormalisation of the model}
\label{sec:renormalisation}
\subsection{Automation of the calculation}
A new model file for the IDM allowing different schemes for the renormalisation of the model, to which we turn shortly, has been added to {\tt SloopS}~\cite{Boudjema_2005, Baro:2007em,Baro:2008bg, Baro:2009na,Boudjema:2011ig,Boudjema:2014gza,Belanger:2016tqb,Belanger:2017rgu,Banerjee:2019luv}. {\tt SloopS}, our automated code for the calculation of tree-level and one-loop observables relies of the bundle of packages based on  {\tt FeynArts}\cite{Hahn:2000kx}, {\tt FormCalc}\cite{Hahn_2016} and {\tt LoopTools}\cite{Hahn:1998yk}, that we refer to as {\tt FFL} for short. A few improvements to {\tt LoopTools}~\cite{Boudjema_2005} have been made over the years. A key component is the generation of the model file (with counterterms and renormalisation conditions) made possible with {\tt LANHEP}~\cite{Semenov:2008jy,Semenov:2014rea} judiciously interfaced with the bundle {\tt FFL}. We have exploited this approach successfully for the SM, the full renormalisation of all sectors of the MSSM~\cite{Baro:2007em, Baro:2008bg, Boudjema:2014gza} and the NMSSM~\cite{Belanger:2017rgu}. The code has been thoroughly checked and allows many tests on the correctness of the results, including tests on the ultraviolet finiteness (for loop calculations) and very importantly, on gauge parameter independence of the results both at tree-level and at loop-level, see~\ref{sec:gauge-fixing} below. The code is optimised also for DM annihilation cross-section such that it inputs directly the result for $\sigma v$, where $\sigma$ is the annihilation cross-section and $v$ is the relative velocity of the annihilating particles. This avoids potential instabilities when specialising to $v \to 0$. 

\subsection{The SM parameters}
\label{sec:SMparam}
The SM part of the IDM is renormalised exactly in the same way as what has become standard practice in one-loop calculations of the electroweak observables~\cite{Belanger:2017rgu}. This calls for an OS scheme whereby the fermion masses as well as the mass of the $W$, $M_W=80.449$ GeV, the $Z$, $M_Z=91.187$ GeV and the Higgs, $M_h=125$ GeV, are taken as input physical masses. For the SM, the tadpole, $T=v (\mu_1^2 -\l_1 v^2)$ (at tree-level) is required to vanish at all orders. The electric charge is defined in the Thomson limit, see~\cite{Baro:2007em}. The light quark ($u,d,s,c$) masses, $m_u=m_d=66$ MeV, $m_s=150$ MeV, $m_c=1.6$ GeV, are taken as effective quark masses that reproduce the SM value of $\alpha^{-1}(M_Z^2) \sim 128.907$. Use of the latter effective coupling can, in many instances,  amount to about $13\%$ correction compared to the use of $\alpha=\alpha(0)=137.036$. Since it is $\alpha(M_Z^2)$ that is used as the effective coupling in the tree-level evaluation of {\tt micrOMEGAs}~\cite{Belanger:2001fz, Belanger:2006is, Belanger:2013oya, Belanger:2018mqt}, we will investigate whether implementing the numerical value $\alpha(M_Z^2)$ instead of $\alpha$ in tree-level cross-sections can account for a significant part or the full one-loop correction. For a portion of the allowed parameter space of the IDM, the relic density is driven almost entirely through the SM Higgs {\it resonance} for which $h \to b \bar b$ is dominant. We will therefore take an effective $b$-quark mass that (at tree-level) reproduces very well the dominant partial width to $b$s and the total SM Higgs width. In fact, for such a scenario, we adapt the renormalisation of the Higgs mass to include its width as well. We give more details~\cite{OurPaper4_2020} on this subtle but important issue when we describe the processes for the DM annihilation that we will be dealing with. The top mass is taken as $m_t=174.3$ GeV.

\subsection{Gauge fixing}
\label{sec:gauge-fixing}
We take a non-linear gauge fixing term~\cite{Boudjema:1995cb, Boudjema_2005, Baro:2007em} that still preserves a $Z_2$ symmetry. We take these gauge fixing terms to be \textit{renormalised}. In particular, the gauge functions involve the physical fields. Although this will not make all Green's functions finite, it is enough to make all $S$-matrix elements finite. With $A_\mu$, the photon field, we write the  gauge-fixing as
\begin{eqnarray}
\mathcal{L}^{GF}&=&-\frac{1}{\xi_{W}}F^{+}F^{-}-\frac{1}{2\xi_{Z}}|F^{Z}|^{2}-\frac{1}{2\xi_{\gamma}}|F^{A_\mu}|^{2}
\, ,
\end{eqnarray}
where
\begin{eqnarray}
F^{+}&=&(\partial_{\mu}-ie\tilde{\alpha}\gamma_{\mu}-ie\frac{c_{W}}{s_{W}}\tilde{\beta}Z_{\mu})W^{\mu +}
+i\xi_{W}\frac{e}{2s_{W}}(v+\tilde{\delta}h+i\tilde{\kappa}G^{0})G^{+} \, ,\nonumber\\
F^{Z}&=&\partial_{\mu}Z^{\mu}
+\xi_{Z}\frac{e}{s_{2W}}(v+\tilde{\epsilon}h)G^{0} \, ,\nonumber \\
F^{A_\mu}&=&\partial_{\mu}A^{\mu} \, .
\end{eqnarray}
The ghost Lagrangian ${{\cal L}}^{Gh}$ is derived through the use of the BRST transformation, see~\cite{Boudjema:1995cb}, which leads to an easy implementation in the automated code {\tt SloopS}. We specialise to the case $\xi_W=\xi_Z=\xi_\gamma=1$ in order that the gauge propagators retain the simple form of the usual Feynman gauge but the non-linear gauge furnishes enough parameters ($\tilde{\alpha},...\tilde{\epsilon}$) on which to carry the gauge parameter dependence of the amplitude for further checks on the correctness of the calculation beside the ultraviolet finiteness tests. We carry gauge parameter independence checks on all cross-sections and decays we calculate with {\tt SloopS} in this series of papers. 

\subsection{On-shell renormalisation for the masses of the IDM scalars}
The renormalisation of the IDM is technically quite straightforward since the new scalars, $A,X$ and $H^\pm$, do not mix, nor do they mix ($1\to 1$ transition) with those of the SM, including the Goldstones, because of the $Z_2$ symmetry. Shifts on the parameters of the potential (which get translated into counterterms, $\delta \l_i$, for the $\l_i$ (and $\mu_2$) as well as shifts on the fields through wave function renormalisation constants ($\delta Z$), are introduced. In the OS scheme, the masses of the physical scalars, $\phi$ ($\phi=X,A,H^\pm$), are defined from the pole position of the one-loop corresponding self-energy $\Sigma_{\phi \phi}(k^2)$, where $k$ is the momentum carried by $\phi$. We also require that the residue at the pole of these particles be properly normalised to unity. The conditions on the counterterms are then
 
\begin{align}
\label{ct_deltam}
 \delta M_\phi^2 &= \text{Re}\Sigma_{\phi\phi}(M_\phi^2),   \\
 \label{ct_deltaZ}
 \delta Z_\phi &= - \text{Re}\left.\frac{\partial \Sigma_{\phi\phi}(k^2)}{\partial
k^2}\right|_{k^2 = M_\phi^2}.
\end{align}\noi

\subsection{$\l_L$: $\overline{{\rm MS}}$ and OS}
\label{sec:Llschmes}
We use the full $h\to XX$ {\em amplitude} to define $\l_L$. At tree-level, the amplitude is given by Equation~\ref{eq:deflambdaL}. It serves to generate (and define) the counterterms for the one-loop amplitude. The latter also includes the contributions of the one-loop diagrams such that the full renormalised one-loop amplitude is momentum dependent. The full one-loop renormalised amplitude (when the threshold is open) for $h(Q^2) \to X(p_1^2) X(p_2^2)$ of the SM Higgs, $h$, with momentum $Q$ to a pair of the DM $X$ with momenta $p_1$ and $p_2$ writes as 
\beqn
{\cal{A}}_{hXX}^{\rm ren}(Q^2,p_1^2,p_2^2)=- \left( v \delta
\l_L + \l_L\delta v+ \l_L v \bigg(\frac{1}{2}\delta Z_h + \delta Z_X \bigg)\right) + {\cal{A}}^{\rm 1PI}_{hXX}(Q^2, p_1^2,p_2^2),
\eeqn
where ${\cal{A}}^{\rm 1PI}_{hXX}(Q^2,p_1^2,p_2^2)$ is the full one-loop 1-particle irreducible vertex. $\delta \l_L$ is the counterterm for $\l_L$ for which we are seeking a renormalisation condition, $\delta Z_h$ is the SM wave function renormalisation and $\delta Z_X$ is the wave function renormalisation for the DM particle, $X$. $\delta v$ is the counterterm for $v$ (defined through $e,M_Z,M_W$ in Equation~\ref{eq:def_sw2}). When the threshold for the Higgs decay to $XX$ is open, we set $Q^2=M_h^2$ and $p_1^2=p_2^2=M_X^2$ and require that the full one-loop correction to this partial width is zero, defining a {\bf gauge invariant} OS counterterm for $\l_L$ as 
\begin{equation}
\label{eq:ctoslL}
 \delta^{{\rm OS}} \l_L= 
\frac{{\cal{A}}^{\rm 1PI}_{hXX}(Q^2=M_h^2,M_X^2,M_X^2)}{v}
-\l_L \Bigg(\frac{\delta v}{v} +\frac{1}{2}\delta Z_h +\delta
Z_X \Bigg) \quad \text{for} \; M_h > 2 M_X.
\end{equation}\noi 

Another gauge invariant but {\em scale dependent scheme} valid even when the decay threshold is closed, is to use a $\overline{{\rm MS}}$ definition where only the (mass independent term) ultraviolet divergent part is kept.

\beqn
\label{eq:renLalMsbar}
 \delta^{\overline{{\rm MS}}} \l_L = \Bigg( \frac{{\cal{A}}^{\rm 1PI}_{hXX}(Q^2,M_X^2,M_X^2)}{v}
-\l_L \Bigg(\frac{\delta v}{v} +\frac{1}{2}\delta Z_h +\delta
Z_H \Bigg)   \Bigg)_\infty   \quad \text{ for any}\; Q^2.
\eeqn

In our code, $\delta^{{\rm OS}} \l_L$ in Equation~\ref{eq:ctoslL} and $\delta^{\overline{{\rm MS}}} \l_L$ in Equation~\ref{eq:renLalMsbar} are extracted exactly by evaluating the amplitude $h \to XX$ and constructing the above equations. 

\subsection{The $\beta$ constant for $\beta_{\l_L}$}
\noi The coefficient of the ultraviolet divergent part is nothing but the one-loop $\beta$ constant for $\l_L$. 

\beqn
\label{deltalLMS}
\delta^{\overline{{\rm MS}}} \l_L =\frac{1}{32 \pi^2} \tilde{\beta}_{\l_L} C_{{\rm UV}}, \quad C_{{\rm UV}}=-\frac{2}{\varepsilon} - 1+\gamma_E-\ln(4\pi),
\eeqn
where $\varepsilon=4-d$ with $d$ being the number of dimensions in dimensional regularisation and $\gamma_E$ is the Euler-Mascheroni constant. Or, keeping $\mu_{\text{dim}}$, the scale introduced by dimensional regularisation (which goes hand in hand with the $\ln(4 \pi)$ term as $\ln(4\pi\mu_{\text{dim}}^2)$), the scale, $Q^2$, dependence through the parameter $\l_L$ writes also as
\beqn
\label{eq:delmulL}
32 \pi^2 \frac{\partial \l_L}{\partial \ln(Q^2)}=-32 \pi^2 \frac{\partial \l_L}{\partial \ln(\mu_{\text{dim}}^2)}=\tilde{\beta}_{\l_L}.
\eeqn

The analytical formulae for the $\beta$ constants of the IDM have been adapted from those of the general 2HDM (2 Higgs Doublet Model)~\cite{Haber:1993an,Ferreira:2009jb} and can be found in~\cite{Goudelis:2013uca, Khan:2015ipa}. For our purposes, we have reformulated them for $\l_L$, taking into account the scalar(s), the gauge (g) and the Yukawa (Y) contribution as
\beqn
\tilde{\beta}_{\l_L} =\tilde{\beta}^{(g)}_{\l_L}+\tilde{\beta}^{(s)}_{\l_L}+\tilde{\beta}^{(Y)}_{\l_L},
\eeqn

with

\beqn
\label{eq:analytical_betas}
\tilde{\beta}^{(g)}_{\l_L}&=&-3\l_L(3 g^2+ g^{\prime 2}) + \frac{3}{4}(3 g^4+ g^{\prime 4} +2 g^2\;g^{\prime 2} ),\nonumber \\
\tilde{\beta}^{(s)}_{\l_L}&=&4\l_L\bigg(\l_L+ 3(\l_1+\l_2)\bigg) -4(\l_1+\l_2)(\l_4+2\l_5)+2(\l_4^2+2\l_4 \l_5+3\l_5^2), \nonumber \\
\tilde{\beta}^{(Y)}_{\l_L}&=&4 \l_L \hspace*{-0.5cm}\sum_{f={\rm all\; fermions}}  \hspace*{-0.6cm}N_C^f \; \frac{m_f^2}{v^2}.
\eeqn

Here, $m_f$ is the  mass of the fermion $f$ and $N_c^f$ is its corresponding colour factor ($3$ for quarks and  $1$ for leptons). For small $\l_L$, the Yukawa contribution is tiny and the largest contribution is from the top-quark. We find excellent agreement (in fact perfect agreement with machine precision) between these analytical formulae and those extracted from our code according to Equations~\ref{eq:renLalMsbar},~\ref{deltalLMS}. An important observation here is that Equation~\ref{eq:analytical_betas} shows that even if $\l_L=0$, a one-loop induced $\l_L$ is generated. Also, while $\l_2$ is a parameter residing solely in the dark sector and is not involved at tree-level in DM annihilation processes to SM particles, it makes its effect felt at one-loop with the conclusion that a large scale variation due to $\l_2$ can be present apart from other contributions at one-loop. In our code, we can freely vary $\mu_{\text{dim}}^2$ to quantify the scheme dependence in an $\overline{{\rm MS}}$ scheme. Obviously in a fully OS scheme, the check on the UV finiteness of the result means that there is no $\mu_{\text{dim}}^2$ dependence. 

\subsection{Scale dependence of the one-loop corrected $h\to XX$ in the $\overline{{\rm MS}}$}
\label{sec:hxxmsbar}
At tree-level, the partial width for Higgs decay to a pair of DM, $X$, is expressed as 
\beqn
\label{eq:pwhinv}
\Gamma_{h\to XX}&=&\frac{\lambda_L^2 v^2}{32 \pi M_h} \sqrt{1-\frac{4 M_X^2}{M_h^2}}, \quad \bigg( M_h  > 2 M_X \bigg),
\eeqn
showing that the parametric dependence on $\l_L$, a quadratic dependence, of this observable is algebraically straightforward. This is the reason that this observable is chosen as an input to define $\l_L$. Naturally, in the OS scheme, this observable receives no correction. Nonetheless, as we will see later, based on present experimental constraints, which means that the Higgs partial decay rate to DM is way too small to be measured, let alone be measured with good precision, one may be forced to rely on an $\overline{{\rm MS}}$ prescription. When $h \to XX$ is closed, the use of the OS scheme is not possible and in this case we have to study the scale uncertainty. 

The purpose of this short interlude is to investigate which scale choice may be considered the best, best in the sense of minimising the one-loop correction. In fact, $\Gamma_{h\to XX}$ provides a good example. One can study this observable at one-loop in the $\overline{{\rm MS}}$ and quantify how it deviates from the tree-level result which is, by construction, the result of the OS scheme. Knowing the $\l_L$ dependence of an observable at tree-level and having at our disposal the $\beta_{\l_L}$ of the model, the scale dependence can be derived analytically allowing to compare the difference between two choices of scale, $\mu_1$ and $\mu_2$.

The $\l_L$ dependence of $\Gamma_{h \to XX}$ is trivial. Indeed, from Equation~\ref{eq:pwhinv}, we have
\beqn
\frac{\delta \Gamma_{h \to XX}}{\Gamma_{h \to XX}}=2 \frac{\delta \l_L}{\l_L},
\eeqn
which through Equation~\ref{eq:delmulL} allows us to relate the one-loop correction at scale $\mu_2$ to that at scale $\mu_1$ 
as 

\beqn
\label{eq:deltahxx}
\frac{\delta \Gamma_{h \to XX}(\bar \mu_2)-\delta \Gamma_{h \to XX}(\bar \mu_1)}{\Gamma_{h \to XX}^{({\rm tree})}}=-\frac{1}{8 \pi^2} \frac{\bll}{\l_L} \ln(\bar\mu_2/\bar \mu_1).   
\eeqn

As expected, an important lesson to always keep in mind is that large $\bll$ values induce large scale variations. For the IDM $\beta_{\l_L}$ depends also on $\l_2$ which can contribute significantly. These considerations are crucial. However, they do not point to the most optimal choice of $\mu$. The most optimal choice of $\mu$ is the one that minimises the full one-loop correction. The latter will involve all scales of the problem. Therefore, to investigate the issue of the optimal scale, we need to consider a few examples calculating the full one-loop correction. 

For the purpose of this exercise, we take a value of $\l_L$ which is not too small in order not to induce unnaturally large relative corrections. At this stage, we do not impose experimental constraints on the parameter space. This is studied in great details, later. Our aim here is to see if there is a trend for an optimal choice of the scale that we could then advocate for other processes driving the relic density. We consider 3 models that differ in the masses of $A$ and $H^\pm$. With $M_h=125$ GeV, we keep $M_X=57$ GeV for these 3 models. These choices generate different hierarchies for the scale of the problem. While $\l_2$ is not needed to calculate the tree-level decay, the value of $\l_2$ is required at one-loop. We consider 3 values of $\l_2$ for each point to gauge the $\l_2$ dependence and we test various values of the scale, $\mu$, in relation with the scales that are involved in the observable at one-loop.  It is important to observe that the masses of the IDM act at two levels in the one-loop result. Re-interpreted in terms of $\l_{i=3,4,5,(L)}$, they contribute directly to the coefficient $\bll$ as $\l_2$ does but unlike $\l_2$, the masses are involved in the arguments of the various one-loop scalar functions.

Our results of the numerical full one-loop corrections are shown in Table~\ref{tab:hxxms}. First of all, Equation~\ref{eq:deltahxx} is in perfect agreement with the numbers given in Table~\ref{tab:hxxms} where we compare, for the same point, the correction between two scales. This is another evidence of the correctness of the implementation of the model and the numerical computation. 

\begin{table}[h]
\centering
\begin{tabular}{|l ||c | c |c|c ||}
\hline
 (Model) \hspace*{0.3cm}$\l_L$;\;\;\;\;$M_A$,$M_{H^\pm}$& $\mu_{{\rm dim}}$&$\l_2=0.$& $\l_2=1$ &$\l_2=2$ \\ 
\hline
\hline
(A) $0.05$;\;\;\;$106,106$&  $M_h/2$ & $0.90$ &$1.48$& $2.05$   \\
$\l_4=\l_5=-0.127,  \l_3=0.305$&   $M_h$ & $-1.9\;10^{-2}$&  -1.36& -2.69 \\
($\bll= 1.03+2.13\l_2$)&  $ 2 M_h$ &-0.94&{\bf -4.18}   &\underline{{\bf -7.44 }}\\
&  $ M_X$ & 1.02 &1.85&2.68 \\
&  $ 2\;M_X$ & 0.10 &-0.98&-2.06\\
&  $ 200$ & -0.64 &-3.28&\underline{{\bf -5.90}} \\
&   $ M_A=M_{H^\pm}$  &$0.2$  & -0.68&-1.56\\ \hline
(B) $0.05$;\;\;\;$138,138$&  $M_h/2$ & $1.73$ &${\bf 4.47}$& {\bf 7.20}   \\
$\l_4=\l_5=-0.252, \l_3=0.555$& $M_h$ & 0.13 & -0.38&-0.90\\
($\bll= 1.78+3.63\l_2$)&  $ 2 M_h$ & -1.46&\underline{{\bf -5.23}}&\underline{{\bf -9.01}} \\
&  $ M_X$ &1.95 & {\bf 5.11}&\underline{ {\bf 8.28}}\\
&  $ 2\;M_X$ & 0.35 &0.26&0.17\\
&  $ 200$ & -0.95 & {\bf -3.67}&\underline{{\bf -6.40}}\\
&   $ M_A=M_{H^\pm}$  & -0.10 & -1.07&-2.06\\
\hline
(C) $0.05$;\;\;\;$170,200$&  $M_h/2$ & {\bf 6.51} & {\bf 15.5}& {\bf 24.4}   \\
$\l_4=\l_5=-0.765, \l_3=1.225$&    $M_h$ & $2.15$ &{\bf 4.87}&{\bf 7.60} \\
($\bll= 4.86+6.94\l_2$) & $ 2 M_h$ & -2.21 &\underline{{\bf -5.71}}&\underline{{\bf -9.21} }\\
&  $ M_X$ & {\bf 7.08} & {\bf 16.9}& {\bf 26.6}\\
&  $ 2\;M_X$ & 2.72 &{\bf 6.28}&{\bf 9.83}\\
&  $ M_{H^\pm}$ & -0.81 & -2.30&-3.80\\
&  $ M_A$ & 0.22 & 0.18&0.15\\\hline
\hline\end{tabular}
\caption{\label{tab:hxxms} {\it One-loop corrections, $\delta \Gamma_{h\to XX}$, to the partial decay width of the SM Higgs ($M_h=125$ GeV) to pairs of DM, $X$, with $M_X=57$ GeV (all values for the widths are given in MeV) in the $\overline{{\rm MS}}$ scheme for different masses of $M_{H^\pm}$ and $M_A$ and $\l_2$, the coupling within the dark sector. The tree-level width of this observable is $5.106$ MeV. All masses (and scales) are in GeV. For each set of parameters defining the IDM model, we give the reconstructed $\l_i$ parameters as well as the one-loop $\tilde{\beta}$. The parameters of the model are only illustrative and do not necessarily pass the experimental constraints on the IDM. Entries in {\bf bold} represent corrections that represent more than $75\%$ of the tree-level value, while those that are also underlined lead to a total width that turns negative.}}
\end{table}

The relevant scales of the problem are all the invariants that are involved in the loop functions that enter the radiative corrections. They include both the external kinematical variables but also the internal masses called in the loop calculations. The $\mu_{{\rm dim}}$ dependence is tracked as $\log(Q_{\Delta}^2/\mu_{{\rm dim}}^2)$. $Q_{\Delta}$ is an effective collective scale which is a combination of the scales involved in the (various) loops entering the loop calculation. As argued at some length in Ref.~\cite{Belanger:2017rgu}, if one  of these scales is (much) larger than the others, it should approximate $Q_{\Delta}$. 

One would expect that the typical scale for this decay is $M_h$. Observe that in our case $M_h \sim 2M_X$. Table~\ref{tab:hxxms} shows that $\mu_{{\rm dim}}=M_h, 2M_X$ is a good choice but only if $M_A<M_h,2 M_X$. In particular for case (C) of Table~\ref{tab:hxxms} where $M_A$ is much larger that $M_h$ and $2M_X$, $M_A$ is by far the best choice for all values of $\l_2$ confirming the general observation made in~\cite{Belanger:2017rgu} in the totally different context of the NMSSM where it was found that the largest scale minimises the correction. 
Keep in mind also that when $M_A$ is much larger than $M_X$, $\beta_{\l_L}$ is larger (since $\l_{3,4,5}$ are larger). It is therefore more important to choose $M_A$ as the optimal $\mu_{{\rm dim}}$. As a rule of thumb we advocate the optimal scale to be $\mu={\rm max}(M_h,M_A)$, $M_h$ is the typical (kinematical) scale of the process while $M_A$ is the typical internal scale. We find that it fares better than $M_{H^\pm}$. Scales that are much smaller or much larger than the optimal scale, for instance $M_X,M_h/2, 2 M_h$, lead to large corrections. On many instances these corrections are so large that the calculations are not reliable. 

We check this conjecture about the optimal scale also in the study of the annihilation cross-section of DM, $XX \to {\rm SM}$. Considering the rather small velocities taking part in the annihilation cross-sections, $2M_X$ will, again represent the typical kinematical scale to be compared again with the internal mass, $M_A$. 


\section{Present Constraints on the parameter space of the IDM}
\label{sec:constraints}
We quantify the effect of the one-loop corrections on the annihilation cross-sections in the IDM. We perform these calculations on some benchmark points of the IDM. The benchmark points need to pass several constraints. While searching for benchmark points that satisfy all theoretical and experimental constraints, we perform a numerical scan where all the constraints that we discuss in this section are implemented numerically without any approximation, by running different codes and routines. Nonetheless, it is instructive to extract the salient features through approximate analytical formulae and some judicious considerations as this helps carry out a more efficient scan. \\
\noi Thorough investigations of the constraints have been performed recently~\cite{Ilnicka:2015jba, Belyaev:2016lok}~\footnote{See also \cite{LopezHonorez:2006gr, Cao:2007rm, Agrawal:2008xz, Lundstrom:2008ai, Andreas:2009hj, Arina:2009um, Dolle:2009ft, Nezri:2009jd, Hambye:2009pw, Miao:2010rg, Gong:2012ri, Gustafsson:2012aj, Swiezewska:2012eh, Wang:2012zv, Goudelis:2013uca, Arhrib:2013ela, Krawczyk:2013jta, Osland:2013sla, Abe:2015rja, Arhrib:2015hoa, Blinov:2015qva, Diaz:2015pyv, Ilnicka:2015jba, Belanger:2015kga, Carmona:2015haa, Queiroz:2015utg, Kanemura:2016sos, Belyaev:2016lok,Eiteneuer:2017hoh, Ilnicka:2018def, Kalinowski:2018ylg, Arcadi:2019lka, Kalinowski:2020rmb}}. While we agree with their findings, we update the constraints in view of some new experimental data on the direct detection of dark matter and searches/analyses at the LHC. Our purpose is not to present new exclusion zones in the parameter space of the model but to make sure that the benchmark points we pick up for our studies at one-loop, pass all the experimental (and theoretical constraints) and are representative of a larger set. As is known, the present precision on the relic density of DM is such that it reduces the parameter space of any model of DM, drastically. However, since all studies impose this constraint based on tree-level calculations of the annihilation cross-sections, we use the tree-level based relic density constraint only as a guide and see how the prediction transforms when a full one-loop calculation is implemented and how the theoretical uncertainties should be taken into account. Many studies have allowed benchmarks where the predicted (tree-level) relic density leads to underabundance, but not (obviously) overabundance. Nonetheless, a prediction of an overabundance with a tree-level calculation may turn into a viable model with a more precise calculation. First of all, let us repeat again that we are taking $X$ as the DM candidate. $X$ is neutral and is taken to be the lightest of the three additional scalars that the IDM provides. From~\ref{eq:livsms} this means that 
\beqn
\label{eq:cl5}
\l_5 < 0. 
\eeqn
Moreover, we take $M_A \leq M_{H^\pm}$, in which case 
\beqn
\label{eq:cl4}
\l_4 \leq \l_5 <0.
\eeqn
The equality ($\l_4=\l_5$) automatically evades indirect electroweak precision measurements by preserving custodial symmetry. Remember also that the existence of the SM Higgs means that 
\beqn
\l_1> 0. 
\eeqn
Usually while studying the viable parameter space of the IDM, the discussion starts with the restriction from the stability and the perturbativity of the potential. We discuss all these constraints but we prefer to start with a constraint that has in the last years become quite stringent. It also delimits a specific coupling and not a combination of a large number of parameters, it concerns the parameter $\l_L$ and its connection to the DM direct detection. $\l_L$ is a key input in our OS renormalisation scheme. 

\subsection{Dark matter direct detection}
The direct detection of dark matter sets a very strong constraint on $\lambda_L/M_X$ for the IDM. In the rather narrow (light) mass range, direct detection proceeds through the SM Higgs exchange with a cross-section given by 
\beqn
\sigma_{\rm SI}^{\rm h} \sim \bigg( \frac{\lambda_L \times 10^{3}}{M_X/100 {\rm GeV}} \bigg)^2\; 8.53 \times 10^{-49} \ {\rm cm}^2.
\eeqn
Assuming that $X$ accounts for all of the DM density on Earth and provides the correct measured relic density as we are assuming, the latest Xenon1T limit~\cite{Aprile:2018dbl} for a DM in the range $50 < M_X < 100$ GeV translates into 
\beqn
\sigma_{\rm SI}^{\rm Xenon1T} < 12 \times 10^{-47} \times (M_X/100 {\rm~GeV}) {\rm cm^2}. 
\eeqn
Applied to our case, we have 
\beqn
\label{eq:llmax}
|\lambda_L| \times 10^{3}< 12 \times (M_X/100{\rm~GeV})^{3/2} &\to& \quad |\l_L|< |\l_L^{\text{max}}|=7 \times 10^{-3} , \;\; \textrm{for} \;\; M_X=70\;\textrm{GeV}, \nonumber \\
& \to& \quad |\l_L|< 5 \times 10^{-3} \;\; \text{for} \;\;M_X=57\;~\text{GeV}. 
\eeqn

This invisible branching fraction is far too small compared to the present limit from the LHC fits~\cite{Aaboud:2018puo, Sirunyan:2019twz, Kraml:2019sis}. Indeed, from  Equation~\ref{eq:pwhinv}
\beqn
\label{eq:hinv}
\text{Br}(h \to XX)&=& \Bigg( 1.183 \; (\lambda_L \times  10^3)^2 \; \sqrt{1-0.64 (M_X({\rm GeV})/50)^2} \Bigg) \; \times 10^{-3}  \\
&=& 2.8 \times 10^{-3} \;\; \text{for} \;\; M_X=57\;\text{GeV} \;,\; \l_L=2.4 \times 10^{-3}.
\eeqn
Considering the very tiny value of the invisible width, we take the total Higgs width to be the SM value of the total width from~\cite{10.1093/ptep/ptaa104, deFlorian:2016spz}
\beqn
\label{eq:HiggsTotalWidth}
\Gamma_h^{\text{SM}}=4.07\pm 0.16 \;\text{MeV}.
\eeqn
If the IDM only provides a fraction of the total relic density of DM, the direct detection cross-section needs to be rescaled. With the density of the IDM written as $\Omega_{{\rm X}}$ and that extracted from the Planck measurements as $\Omega_{{\rm DM}}^{{\rm Planck}}$, the limits above are transformed into 
\beqn
\l_L < \l_L^{\text{max}} \sqrt{\frac{\Omega_{\text{DM}}^{\text{ Planck}}}{\Omega_{\text{X}}}},
\eeqn
where $\l_L^{\text{max}}$ is derived from Equation~\ref{eq:llmax}. We take the view that if the IDM is to be considered as a model for DM, it should provide at least $50\%$ of the total DM. This requirement does not significantly change the limit on $\l_L$ from direct detection constraints (a modest factor of $\sqrt{2}$ is possible) which is restricted to be rather small, $\l_L<0.01$. 

\subsection{Electroweak Precision Observables (EWPO)}

The custodial $SU(2)$ symmetry breaking parameter, $T$~\cite{Peskin:1991sw}, restricts mass splitting between the scalars of the IDM. In the limit $M_X \ll M_A,M_H$, we require $M_A\sim M_H$. Indeed, 

\begin{equation}
 \Delta T \simeq \frac{1}{24 \pi^2 \alpha v^2}M_A  \left(M_{H^\pm}-M_A\right) \sim 0.05 \frac{M_A}{500\; \textrm{GeV}}\frac{\Delta M}{10\; \textrm{GeV}},
\end{equation}

\noi is combined with the $S$~\cite{Peskin:1991sw} parameter which gives the weaker constraint
\beqn
\Delta S \simeq -\frac{5}{72 \pi}.
\eeqn

\noi The full expressions for the $S,T$ contribution in the IDM can be found in Ref.~\cite{Barbieri:2006dq}.

\noi In the code, we impose~\cite{Baak_2012}
\beqn
S = 0.06 \pm 0.09, \quad  T = 0.10 \pm 0.08, \text{with a correlation coefficient of} +0.89. 
\eeqn

\subsection{Stability of the potential}

As with $\l_1 > 0$, we need to have $\l_2 > 0$. Equations~\ref{eq:cl5} and~\ref{eq:cl4} are sufficient to guarantee the vacuum to be neutral ($\l_4-|\l_5|<0$). The other constraints on the potential (see for example~\cite{Belyaev:2016lok}) are easily satisfied if one takes into account that direct detection requires a very small $\l_L$. The conditions expressed as a function of $\l_L$ require for instance  
\beqn
2\sqrt{\l_1\l_2}+\lambda_L > 0,
\eeqn
which is satisfied if we take $\l_L > 0$. In our case, the condition $2\sqrt{\l_1\l_2}+\l_3=2\sqrt{\l_1 \l_2} +\l_L-\l_4-\l_5>0$ would be redundant.

Moreover, with very small $\lambda_L$ and not vanishingly small $\l_2$, the parameter, $R=\frac{\l_L}{2 \sqrt{\l_1 \l_2}}$, satisfies $R \ll 1$ and we have that the minimum of the potential, which is the trivial one with the inert minimum being the deepest~\cite{Belyaev:2016lok}. Requiring 
\beqn
\label{eq:l2min}
\l_2 > 0.01,
\eeqn
satisfies these conditions. An upper limit on $\l_2$ ($\l_2$ enters at one-loop order in the relic density calculation) is derived by considering unitarity to which we now turn to. 

\subsection{Tree-level unitarity}

Unitarity  constraints (that are found to be stronger than the perturbativity constraints) are also imposed, see~\cite{Arhrib:2012ia}. These have been studied by considering the eigenvalues of the full set of all scattering processes involving all the scalars in the theory.  They allow us to set an upper limit on the masses of the scalars. For $M_{H^\pm} \gg M_X$, these constraints simplify if we impose the $T$ parameter constraints, $M_A=M_{H^\pm}$ (or $\l_4=\l_5=-\l$). They can be decomposed into subsets. The $\l_1,\l_2$ independent limits translate into 
\beqn
\label{eq:MAmax}
M_A \leq \sqrt{\frac{8 \pi}{3}} v \sim 720 \; \text{GeV}.
\eeqn

Stronger limits apply if we allow for non-negligible values of $\l_2$. In the approximation, $\l_4=\l_5 $ and with $\l_L \sim 0$, we have the additional constraint
\beqn
\label{eq:l2MAmax}
(\l_1+\l_2) +\sqrt{(\l_1+\l_2)^2+\l^2} < \frac{8\pi}{3}, 
\eeqn
which, for small $\l_2$, gives the limit in Eq.~\ref{eq:MAmax}. For $\l_2=2 \; (1)$  values as high as $M_A=600 \; (660)$ GeV are within the perturbative tree-level regime. Tree-level unitarity allows much larger values of $\l_2$ than 2, especially for not too large values $M_{H^\pm} \sim M_A$ as implied from equation~\ref{eq:l2MAmax}. We have however restricted our analyses to $\l_2=0.01,1,2$. 

\subsection{Collider limits}

\subsubsection{Direct searches at LEP}

The LEP constraints can be easily evaded if the masses of the inert scalars are above the threshold for LEPII production. In our benchmarks, this applies to $H^+ H^-$ pair production. For $XA$ associated production at LEPII, avoiding the constraints is possible either because the cross-section is much reduced (close to threshold) and/or because the signature is such that these particles go undetected. Reinterpretation in terms of the IDM of LEPII data for searches of charginos~\cite{Pierce:2007ut} and neutralinos~\cite{Lundstrom:2008ai} has been done. These constraints can be summarised simply as $M_A,M_{H^\pm} > 110$ GeV independently of the $X$ mass (as long as $M_X < M_{A,H^\pm}$). There is however a caveat when $\Delta M_{AX}=M_A-M_X< 8 \text{GeV}$ that leads to too soft leptons that invalidate the LEP searches. This recently discovered small region~\cite{Datta:2016nfz, Belyaev:2016lok} allows efficient $XA$ co-annihilation into a fermion pair. We study this region and look at the impact of the loop corrections on the relic density in Ref.~\cite{OurPaper2_2020}. 

\subsubsection{Direct searches at the LHC}
The searches (and limits on the parameter space) are based primarily on the Drell-Yan like associated production of the scalars. The signatures are classified according to the number, $\ell$, of \underline{charged} leptons, missing energy and possibly jets: 
\beqn
\label{HpX}
H^\pm X &\to &W^\pm XX \to \ell^\pm +\slashed{E}_T, \quad \ell=1 \\
\label{AX}
A X &\to& (Z)X X \to \ell^+\ell^- +\slashed{E}_T, \quad \ell=2 \; \quad (\text{mono-Z})\\
\label{HpA}
H^\pm A &\to& (W^\pm) (Z) XX \to \ell^{'\pm} \ell^+ \ell^- +\slashed{E}_T, \quad \ell=3 \\
\label{HpHm}
H^+ H^- & \to& (W^+) (W^-) XX \to \ell^+ \ell^{'-} +\slashed{E}_T, \quad \ell=2 ,
\eeqn
where the $Z,W^\pm$ may or may not be on-shell. The monojet signature ($XXj, XAj$) is relevant only when $\l_L$ is not too small ($XXj$) or when $M_A-M_X$ is very small~\cite{Belyaev:2016lok}. The contribution from vector boson fusion is negligible~\cite{Belyaev:2016lok}. Higher values of $\ell=4,5$ can be envisaged depending on the mass difference between $M_A$ and $M_{H^\pm}$ giving more leptons through cascade decays~\cite{Datta:2016nfz}.

\noi The most studied~\cite{Cao:2007rm, Dolle_2010} scenario is the dilepton scenario in Eq.~\ref{AX}. Reinterpretation of the ATLAS and CMS 8 TeV data concerning searches of charginos, neutralinos and sleptons, to the case of the IDM was conducted in Ref.~\cite{Belanger:2015kga}. They are essentially based on the dilepton signature of the IDM, Eq.~\ref{AX}, with the important caveat that dileptons from the $Z$-boson decay are vetoed by the CMS/ATLAS cut (|$m_{\ell\ell}-M_Z|>10$ GeV). Combined with the LEP data, one learns~\cite{Belanger:2015kga, Ilnicka:2015jba, Belyaev:2016lok} that one should impose~\footnote{This lower limit depends slightly on the masses of the other scalars. For us, it rests a very good starting point to find viable IDM DM candidates.} 
\beqn
 M_X> 45 \text{GeV}.
 \eeqn
 
\noi Simulations for all process~\ref{HpX}-~\ref{HpHm} were studied for some benchmark points in~\cite{Datta:2016nfz} while a full simulation was performed for~\ref{AX} in Ref.~\cite{Belyaev:2016lok}. The caveat, |$m_{\ell\ell}-M_Z| > 10$ GeV, we alluded to earlier, has a crucial impact on the searches, existing and forthcoming. This cut is meant to reduce the large SM background in events containing a (on-shell) $Z$-boson. Yet, for a large part of the IDM parameter space, $\Delta M_{AX}>M_Z$, where $A \to X Z$ proceeds with an on-shell $Z$. Therefore, large values of $M_A$ not only give smaller cross-sections but also a very large fraction of the yield of leptons are cut. Thus, the prospects for future discovery are slim. 

Very recently, ATLAS has also given limits on the masses of the supersymmetric charginos and neutralinos in the pure (simplified) wino and (non-degenerate) higgsino limits of the MSSM (minimal supersymmetric model)~\cite{LHCXSWG} based on data from the LHC at $\sqrt{s}=13$ TeV. In these special manifestations of the MSSM (heavy mass sfermions), the production mechanisms and the signatures at the LHC are the same as those of the IDM. The only (notable) difference is the spin of the produced particles in these models. Therefore, we also take into account the ATLAS exclusion zones observing that the corresponding limits on the IDM are more relaxed because of the scalar nature of the IDM particles resulting in much smaller cross-sections. In Table~\ref{tab:bpfb}, we therefore also provide the corresponding cross-sections for the electroweak production of the IDM alongside those of the wino and higgsino production processes. Moreover, we also carry a recasting analysis based on {\tt MadAnalysis 5}~\cite{Dumont:2014tja, Conte:2018vmg, Araz:2020lnp} on some of the benchmarks points. A more extensive recast of the IDM will be studied independently of this series of papers. We check few of the benchmarks against some of the existing searches. In particular, we use the $2\ell+\slashed{E}_T$~\cite{Aad:2019vnb, DVN/EA4S4D_2020} from electroweakinos/slepton pair production, the $2\ell + \slashed{E}_T, 3\ell + \slashed{E}_T$~\cite{Sirunyan:2017lae, electroweakinos} from the chargino-neutralino pair production, the $2\ell+\slashed{E}_T$~\cite{Sirunyan:2017onm, monoZ} from the mono-$Z$ search and the $\ell + \slashed{E}_T$~\cite{Aad:2019wvl, DVN/GLWLTF_2020} from the $W^\prime$ search. We must mention here that the topologies considered in the latter two analyses are not the same as ours and detailed simulations including relevant cuts are necessary for a proper study.  However, the aforementioned analyses, especially the ones with the chargino pair or the chargino-neutralino productions, show that our benchmark points are safe. We also use {\tt MadAnalysis 5}'s extrapolation code~\cite{Araz:2019otb} to show that the checked benchmark points are also allowed at the high luminosity run of the LHC (HL-LHC) with an integrated luminosity of $\mathcal{L}=3$ ${\rm ab}^{-1}$.

\subsection{$\mu_{\gamma \gamma}$}

We also check for a possible constraint from the signal strength in the diphoton decay of the SM Higgs,
\beqn
\mu_{\gamma \gamma}=\frac{\text{Br}_{\text{IDM}}(h\to \gamma \gamma)}{\text{Br}_{\text{SM}}(h\to \gamma\gamma)}.
\eeqn
With \cite{10.1093/ptep/ptaa104, deFlorian:2016spz}
\beqn
\text{Br}_{\text{SM}}(h\to \gamma \gamma)=(2.27 \pm 0.05)\;10^{-3}.
\eeqn

The ATLAS and CMS limits for $\mu_{\gamma \gamma}$ are given in Refs.~\cite{PhysRevD.98.052005} and~\cite{Sirunyan_2018} respectively. We use the standard combination of signal strengths and uncertainties (see~\cite{Banerjee:2016vrp}) and impose 
\beqn
\mu_{\gamma \gamma} = 1.04 \pm 0.1.
\eeqn
The IDM contribution has been worked out in~\cite{Swiezewska:2012eh}. It turns out, see Table~\ref{tab:bpfb}, that the present signal strength constraint on the IDM is not significant. Nonetheless, we list the values of $\mu_{\gamma \gamma}$ for the benchmark points should future LHC analyses improve the limits. 

\begin{table}[htbp]
\centering
\tiny
\begin{tabular}{|c|c |c ||c|c|| c |c |c |c| c|c|c|c|}
\cline{2-13}
\multicolumn{1}{c|}{} & \multicolumn{2}{c||}{Res. Higgs}& \multicolumn{2}{c||}{Co-ann}
& \multicolumn{8}{c|}{$XX \to WW^\star,ZZ^\star$}\\
\cline{2-13}
\multicolumn{1}{c|}{} & P57& P59&P58&P60& A&B&C&D&E&F&G&H \\
\hline
$M_X$  &  57 & 59& 58&60& 70 &70  & 70& 70& 72& 72 &72&70\\ 
\hline 
$\l_L\times 10^3$  & 2.4&1.0&0.0&0.0& 5.0  &5.0  & 4.7&4.7 &0.5  & 3.8&0.1&7.0\\ \hline
$M_A$         &113 &113& 66&68& 170  &130 &  360& 571&165 & 138 &158&250\\  \hline
$M_{H^\pm}$& 123 &123& 110&150&   200&240 & 360 &571 &165 &138&158&250\\  \hline \hline
   \multicolumn{13}{|c|}{$\Omega h^2$} \\ \cline{2-13}
   \hline
$\alpha(0)$ &  0.113 & 0.108& 0.113&0.116& 0.156 &0.153  &0.146 &0.142&0.119 & 0.119&0.121&0.142\\  \cline{2-12}    
   $\alpha(M_Z^2)$ & 0.118 &0.113&  0.101&0.103& 0.130 &0.128  &0.121 &0.119 &0.099 & 0.099&0.101&0.119\\  \cline{2-13}    
    $\alpha(M_Z^2),\l_L=0$ & 1.97 & 1.97& 0.101&0.103& 0.146 &0.143  &0.135 &0.132 &0.100 & 0.107&0.102&0.140\\ \hline
 $\Omega_{WW^\star}(\%)$      &22 &24 &5&9& 90 &90 &  88& 88& 89 & 88& 88&88\\ \cline{2-13}
$\Omega_{ZZ^\star}(\%)$        &- &-& -& - &10&  10& 12& 12& 11& 12 &12&12\\  \cline{2-13}  
$\Omega_{bb}(\%)$        &  58 & 57&-& - & -& -&- & -& -& -& -&-\\ \cline{2-13}
$\Omega_{gg}(\%)$ &  8 & 7&- &-& -& -&- &- & - & -& -&-\\ \cline{2-13}
 $\Omega_{AX\to f\bar f}(\%)$ &  - & -&95&91 & -& -&- &- & -& -& -&-\\ \cline{2-13}
 \hline
 \hline $\mu_{\gamma \gamma}$ & 1.022 & 1.024 & 1.026 & 1.021 & 1.021 & 1.018 & 1.015 & 1.014 & 1.026 & 1.032 & 1.028 & 1.018 \\  
 \hline \hline
   \multicolumn{13}{|c|}{LHC13 cross-sections} \\ \cline{2-13}
 \hline 
 $A\to ZX$ open    & no     & no     & no      & no     & yes    & no     & yes    & yes    & yes    & no     & yes     & yes\\   
 $H^\pm \to W^\pm X$ open &no&no&no&yes&yes &yes&yes &yes &yes  &no &yes  &yes\\  \hline
 $pp \to AX$       & 364 & 352 & 1475    & 1277   & 94  & 198 & 8   & 1   & 100  & 163 & 113  & 29\\  
higgsino           & 4048   & 3894   & 18310   & 15430  & 1019   & 2184   & 86  & 15  & 1088   & 1800   & 1233    & 305 \\
\hline 
$pp \to H^\pm X$   & 497 & 481 & 659  & 272 & 104 & 59  & 15  & 3   & 178 & 287 & 201  & 52\\  
higgsino           & 5366   & 5199   & 7196    & 2909   & 1123   & 637 & 161 & 28  & 1931   & 3140   & 2177    & 557 \\
\hline
$pp \to H^\pm A$   & 200 & 200 & 566  & 244 & 37  & 35  & 3   & 0.3   & 58  & 112 & 68  & 12\\  
wino               & 8890   & 8890   & 24850   & 10550  & 1756   & 1634   & 138 & 19  & 2666   & 5076   & 3122    & 579\\
higgsino           & 2222   & 2222   & 6201    & 2642   & 438 & 409 & 35  & 4.6   & 669 & 1274   & 781  & 144 \\ \hline
$pp \to H^+ H^-$   & 97  & 97  & 144  & 48  & 17 & 9   & 2   & 0.4   & 34  & 64  & 39   & 7.3\\  
wino               

& 3814   & 3814   & 5704    & 1860   & 650    & 328 & 66  & 9   & 1318   & 2523   & 1539    & 280\\
higgsino           & 1045   & 1045   & 1537    & 521 & 186 & 94  & 19  & 3   & 372 & 699 & 434  & 81 \\ \hline
\hline 
\hline
 \end{tabular}
 \normalsize
 \caption{\label{tab:bpfb}{\it Benchmarks points that pass the constraints discussed in Section~\ref{sec:constraints}. All masses are in GeV. The benchmarks are divided into three classes: 1) the Higgs resonance region, 2) the small co-annihilation region with $\Delta M=M_A-M_X$ allowed by LEPII and 3) the less tuned annihilation class for masses of a DM around $M_X=70$ GeV. For the relic density, we use \texttt{micrOMEGAs} {\bf 5.07} for which  the SM parameters are given in Section~\ref{sec:SMparam}. The relic density is calculated both with an OS $\alpha$ in the Thomson limit and with the effective $\alpha(M_Z^2)$. To weigh the dependence of $\l_L$, we also give the result for $\l_L=0$. The relative contribution of the most important annihilation/co-annihilation (more than $5\%$) cross-sections to the relic density are given: $XX\to W^+W^-, ZZ, b\bar b, gg$ and $AX\to f \bar f$. These relative contributions do not depend much on the value of the $\alpha$ used for the relic density calculation. For the relative contributions, we therefore show only those calculated with $\alpha(0)$. In the particular case of point P57, the relative contributions are changed drastically when $\l_L=0$ (they become $90\%$ into $WW^\star$ and $10\%$ into $ZZ^\star$) since the Higgs mediated $b\bar b$ final state is eliminated with $\l_L=0$. We do not show, in this table, either the values of the $S,T,U$ parameters or of the direct detection. The latter can be trivially rescaled, see Section~\ref{sec:constraints} for the theoretical constraints of unitarity which we carefully check. For each point, we do however give the values of the LHC ($\sqrt{s}=13$TeV with NN23LO1 parton distribution function, pdf, as implemented in {\tt MadGraph5\_aMC\@NLO}~\cite{Alwall_2014}) cross-sections (in fb) for the electroweak production of a pair of the new scalars, should future LHC analyses update the exclusion regions. Alongside these IDM cross-sections, we also list the wino and higgsino cross-sections that the LHC collaborations have studied (simplified wino and non-degenerate higgsino)~\cite{LHCXSWG} and may update in the future. Note that the phenomenology of point A at a high luminosity LHC is considered in~\cite{Belyaev:2016lok} and that of point B in~\cite{Datta:2016nfz}. The LHC cross-sections are computed at LO. NLO results for the cross-sections at the LHC should be scaled by a factor of $30\%$. While the NLO corrections in the case of supersymmetry have been computed, we expect the same corrections for the IDM, since these are DY-like processes and the corrections are essentially initial-state QCD corrections. We also show the values of the Higgs to diphoton signal strength that includes the IDM contribution, should this observable be better constrained in future LHC analyses.}}
\end{table}

\subsection{Relic density}

The present limit~\cite{Ade:2015xua} on the relic density abundance is~\footnote{We quote the so-called TT (Planck high multipole TT likelihood) + lowP (low multipole polarisation data). Adding {\it external, +``lensing",$\cdots$} data does not change the limit appreciably, especially in view of the larger theoretical uncertainty that we include.}
\begin{equation}
\label{omhexp}
 \Omegah = 0.1197 \pm 0.0022.
\end{equation}
\noi The {\em experimental} uncertainty is less than $2\%$. This is the main reason it is extremely important that the theoretical prediction be as precise as possible. Once the cosmological model has been set up, in our case the thermal freeze-out assumption, the major uncertainty is the evaluation of the annihilation cross-sections. These cross-sections must be evaluated beyond the tree-level approximation. This is the aim of this series of papers for the IDM model. However, to select a few points that pass all other constraints and weigh in the importance of the radiative corrections, we have to be content with a prediction of the relic density based on tree-level cross-sections. We use {\tt micrOMEGAs 5.0.7}~\cite{Belanger:2001fz, Belanger:2004yn, Belanger:2006is, Belanger:2013oya, Belanger:2018mqt} for computing the relic densities. Some considerations beyond tree-level are taken into account in {\tt micrOMEGAs}; in particular the Higgs couplings to the SM particles. The code also uses a running for the electromagnetic coupling with an effective coupling estimated at $\alpha(M_Z)\sim 1/128.907$ instead of the OS coupling in the Thomson limit, \textit{i.e.}, $\alpha=/137.036$. For many annihilation cross-sections, this change of $\alpha$ amounts to an overall rescaling that brings a correction of about $13\%$ in $2 \to 2$ annihilation processes. However, in many scenarios and models beyond the SM, this rescaling does not account for the full one-loop correction~\cite{Baro:2007em, Baro:2009na, Boudjema:2011ig, Banerjee:2016vrp, Banerjee:2019luv}. It is therefore wise when we select benchmarks points, using {\tt micrOMEGAs} for the relic density constraint, to allow for a theoretical uncertainty of $20\%$. We will therefore keep points that satisfy  
\beqn
0.096< \Omegah < 0.144. 
\eeqn
Moreover, we quote the value of the relic density using {\tt micrOMEGAs} with both an effective $\alpha(M_Z)$ and $\alpha$. Once we compute the full one-loop corrections, we comment on the use of the effective electromagnetic coupling for each of the benchmark points that we study at one-loop. When running {\tt micrOMEGAs}, we also let the code check explicitly for the direct detection and possible indirect detection constraints.


\section{The different channels contributing to the relic density for $55< M_X <75$ GeV }
\label{sec:summary_channels}
\subsection{Characteristics of the benchmarks points}
Keeping all the constraints that we have introduced so far, we do of course recover the general conclusions of a scan of the IDM with the relic density constraint calculated with {\tt micrOMEGAs}. Namely, apart from a region with the heavy scalar, $M_X>500$ GeV, scenarios, we investigated in~\cite{Banerjee:2019luv}, the low mass region range lies in the range  $55~{\rm GeV} <M_X < 75~{\rm GeV}$.  

Within this relatively small range, 3 different mechanisms are at play
\begin{itemize}
\item {\bf The co-annihilation region}\\
For $M_X \sim M_A$ (within $M_X -M_A<8$ GeV set by the LEPII constraint), extremely efficient {\it co-annihilation} $XA \to f\bar f$ takes place. It is so efficient that the Boltzmann factor is needed to reduce its contribution. Therefore, $\Delta M_{AX}$ should be as large as possible. This is the reason this scenario occurs at the very edge of the limit allowed by LEPII {\it i.e.,} $M_A \sim M_X+8\;$GeV.  For $M_X\sim 55$ GeV, $XX$ annihilations are still away from the $h$ peak and the onset of $XX \to W f \bar f^\prime$, $XA\to f \bar f$ is the principal channel. In any case, for the smallest $M_X$ masses, one still needs to reduce annihilation to the SM Higgs by setting $\l_L\sim 0$. Although we find this scenario extremely fine tuned, we find it a good example for the effect of the loop corrections in a certain limit. Indeed, in this limit the tree-level process is governed totally by the gauge coupling. The one-loop corrections to $AX \to Z \to f \bar f$ are technically the easiest to consider. A detailed investigation is conducted in Ref.~\cite{OurPaper2_2020}.
\item {\bf The SM Higgs resonance} \\
For $M_X \sim M_h/2$, efficient annihilation is possible through the very narrow Higgs resonance. Again, this also fine-tuned. Nonetheless, the annihilation does require non-zero $hXX$, $\l_L$, coupling. The one-loop corrections here will have to deal with the implementation of the width at one-loop, a non trivial problem. We leave all technicalities and discussion on the results in Re.~\cite{OurPaper4_2020}.
\item {\bf The annihilation $XX \to W W^\star, ZZ^\star$}\\ 
For $M_X<80$ GeV, these annihilations occur with large enough rates even though we are below the $WW$ threshold. They become too efficient when this $WW$ threshold is crossed. Past about $M_X \sim 73$ GeV, the annihilations become too large, depleting the dark matter density. Technically, the one-loop corrections here are quite challenging, requiring the computation of $2\to 3$ processes at one-loop. We leave these calculations and their impact on the relic density to a separate paper~\cite{OurPaper3_2020}. It is not possible to attempt a $ 2\to 2$, $XX \to W f \bar f^\prime$, with an off-shell $W$, and integrate over Breit-Wigner distribution without getting into trouble with gauge invariance. Nonetheless, since $WW$ and $ZZ$ annihilation cross-sections are a backbone to the $XX\to V f\bar f$~\cite{OurPaper3_2020}, we study these $2 \to 2, XX \to W^+ W^-, ZZ$ on-shell processes, see section~\ref{sec:XXVVwarmup}, before we embark on the full $2 \to 3$ processes. This investigation helps unravel some characteristics of the corrections and more importantly reveals that one-loop perturbativity may break down for some choices of the parameters that tree-level unitarity allows. This study above the $WW$ and $ZZ$ threshold will help select further among the benchmark points. 
\end{itemize}

\noi These benchmark points based on the constraints we reviewed so far are listed in Table~\ref{tab:bpfb}. Most of the regimes are, at tree-level, driven by the gauge coupling but the contribution of $\l_L$ is not negligible. The value of $\l_L$ plays a major role in the derivation of the cross-section for the Higgs resonance scenario since this mechanism calls for the $h XX$ coupling. The co-annihilation region requires  $\l_L \simeq 0$ and is therefore driven by the gauge coupling. While the (non-resonant) annihilation region, $XX \to W f \bar f^\prime$ and $XX \to Z f \bar f$, is dominated by the gauge coupling, the $\l_L$ contribution is not negligible at all. First, it occurs because there is a contamination from the SM Higgs exchange but also because the production of longitudinal vector bosons is $\l_L$ dependent. This $\l_L$ dependence can be seen in the $XXGG$ and $XXG^+G^-$ quartic couplings which once the masses of the scalars are fixed ($\l_{4,5}$ fixed) the coupling is sensitive to $\l_L$. Table~\ref{tab:bpfb} gives the value of the relic density when $\l_L$ is switched off. Because of the importance of this coupling and the fact that we will, in some cases, rely on an $\overline{MS}$ scheme for its renormalisation, we give in Table~\ref{tab:livaluesBP} the value of the $\beta$ constant of $\l_L$, $\bll$, for each benchmark point. Observe the very large values we obtain for this constant for points E and D that are associated with large values of the heavy scalars. 

\begin{table}[h]
\centering
\begin{tabular}{|c ||c |c| c |c ||}
\hline
Model  &     $\l_4$         & $\l_5$ &$\l_3$ &$\beta_{\l_L}$  \\
\hline 
P57&$ -0.227579$&$ -0.152146$&$0.382125$&$1.12312 + 2.15628 \l_2$   \\
P59&$ -0.223871$&$ -0.148438$&$0.373309$&$1.09738 + 2.09499 \l_2$   \\
\hline
P58&$ -0.263378$&$ -0.015854$&$0.279232$&$0.77300 + 1.18034 \l_2$   \\
P60&$ -0.587742$&$ -0.016365$&$0.604108$&$1.50953 + 2.48189 \l_2$   \\  
\hline
 A&$   -0.738354$&$ -0.383561$&$ 1.12691$&$ 4.34278 + 6.0819\l_2$  \\ 
 B&$ -1.49269$&$ -0.19178$&$ 1.68947$&$ 7.2441 + 7.565 \l_2$   \\ 
 C&$   -1.99292$&$ -1.99292$&$ 3.99053$&$ 51.1306 + 23.9714 \l_2$  \\ 
 D&$  -5.13238$&$ -5.13238$&$ 10.2695$&$ 324.269 + 61.6449 \l_2$   \\
 E&$  -0.352252$&$ -0.352252$&$ 0.705005$&$ 2.48707 + 4.23303 \l_2$   \\
 F&$-0.221506$&$ -0.221506$&$ 0.446813$&$ 1.40176 + 2.70368 \l_2$  \\ 
 G&$   -0.316118$&$ -0.316118$&$ 0.632336$&$ 2.14181 + 3.79461 \l_2$  \\
 H&$   -0.920545$&$ -0.920545$&$ 1.84809$&$ 12.0397 + 11.1305 \l_2$  \\ 
 \hline
 \end{tabular}
 \caption{\label{tab:livaluesBP}Values of the underlying $\l_{3,4,5}$ parameters and the corresponding $\beta$ functions for $\l_L$, $\tilde{\beta}_{\l_L}$, for the benchmarks points defined in Table~\ref{tab:bpfb}.}
\end{table}

\subsection{Cross-sections and velocity dependence}
\begin{figure}[bp]
 \begin{center}
\includegraphics[scale=0.5]{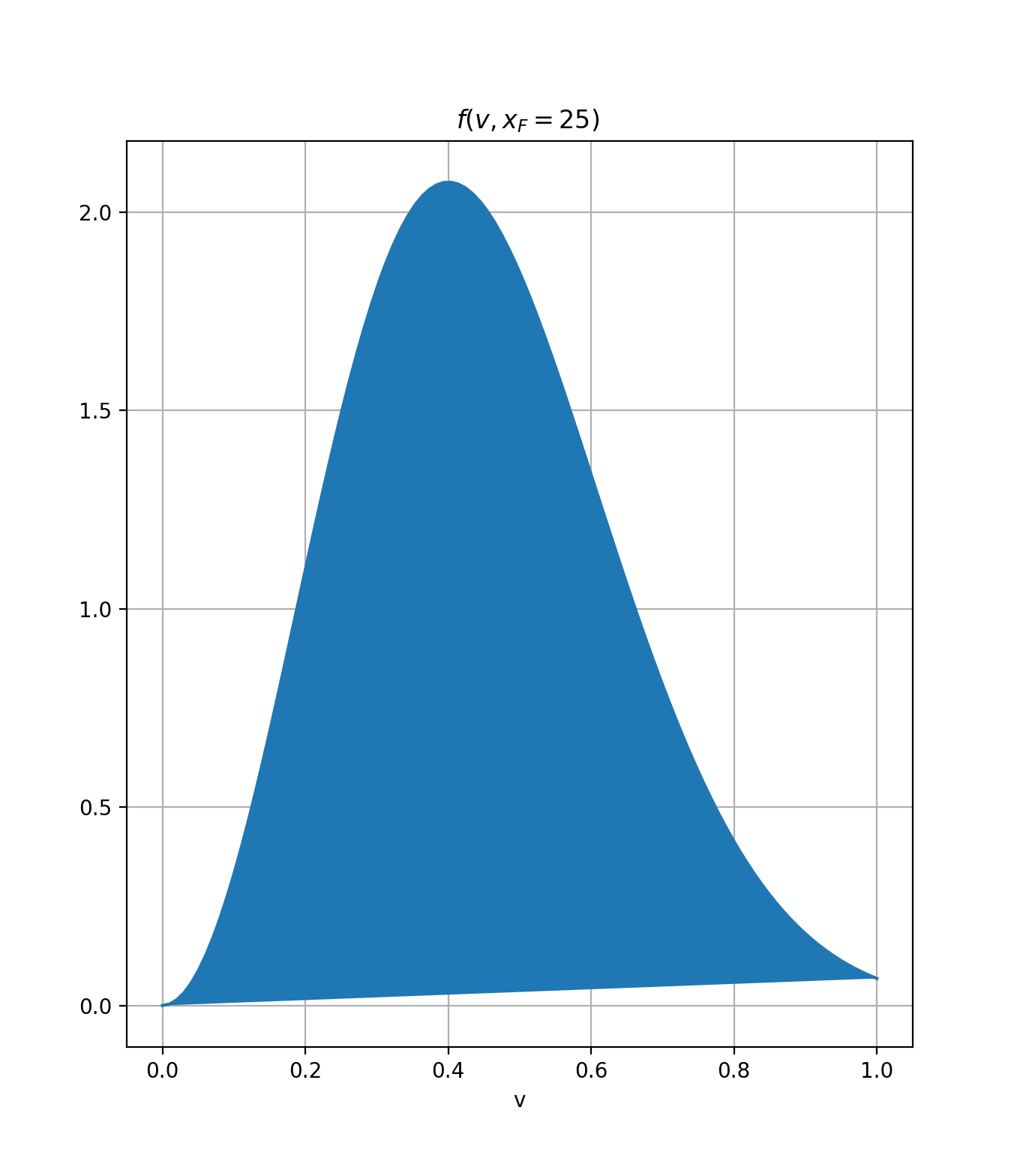}
\caption{\it \label{fig:bellshapev} The normalised velocity distribution $f(v,x_F=25)$ as a function of $v$.}
\end{center}
\end{figure}
In order to convert the cross-sections into their contribution to the relic abundance, a thermodynamical/cosmological model needs to be implemented. In the case of freeze-out that we will be working with, a thermal average and a convolution of the cross-sections over relative velocities of the annihilating particles is performed. The velocity distribution is important since if the cross-sections are large in regions where the weight of the velocity distribution is negligible, the contribution of the process to the relic density is very much reduced. In case of annihilations, $v$ determines the centre of mass total energy, or the kinematical invariant $s$, as 
\beqn
\label{eq:relv}
v=2\sqrt{1-\frac{4 M_X^2}{s}} \quad \text{or} \quad \sqrt{s}=2 M_X \frac{1}{\sqrt{1-\frac{v^2}{4}}}.
\eeqn 
Although we will interface all of our cross-section to {\tt micrOMEGAs} for the velocity and temperature averaging for a full numerical integration, it is instructive to gain an analytical understanding. The velocity averaged cross-section, at a temperature $T$ for annihilating DM particles of mass $M_X$ can be approximated as 
\beqn
\label{eq:vel_fct_relv}
<\sigma v > =\int_0^{\infty} dv f(v, x) (\sigma v); \quad f(v, x) =\frac{x^{3/2}}{2\sqrt{\pi}} v^2 e ^{-x v^2/4}, \quad x=M_X/T\;, (x>1).
\eeqn
Figure~\ref{fig:bellshapev} shows the weight function of the velocity distribution at the freeze-out temperature, $T_f$. As a general rule, freeze-out occurs at $x_f=M_X/T_f \sim 20-25$. This helps understand which range in the relative velocity, $v$, we need to generate most of the annihilation cross-sections. This is particularly useful for the calculations at one-loop cross-sections where there is no need to compute the (involved) one-loop cross-sections for too many values of $v$ where these cross-sections are small. Observe that a typical $v$ where we must get the cross-section right is $v \sim 0.4$ (see figure~\ref{fig:bellshapev}) and that for $v>0.8$ the cross-sections are weighted down. As we will see, there may be exceptions to this general rule~\cite{OurPaper4_2020}. In particular, if the cross-section is dominated by a peak (in our case the Higgs resonance region) that may occur toward the higher end of the distribution function and also, to a lesser extent, if the cross-section grows because of the opening up of a threshold. In our analyses, for the most general case, we compute the cross-sections for values of $v$ as high as $v=0.9$ (or even $v \sim 1$). Going to such high values of $v$ makes sense only if the cross-section peaks dramatically as is the case of the resonance through the very narrow SM Higgs. Indeed, for the point $M_X=57$ GeV, $v=0.82$ corresponds to $\sqrt{s} \simeq M_h=125$ GeV. Observe that for $M_X=70,72$ GeV, $v=0.8$ does not permit the production of both $W$s on-shell but only a $WW^\star$ and therefore we must consider $Wf  \bar f$. 

In the case of co-annihilation, the Boltzmann factor is a very penalising factor. In our case, this concerns $AX$ co-annihilation when the mass difference between these two particles is small. The effective relative weight (to the annihilation $XX$ rate) is 
\beqn
\label{eq:relv_coann}
\tilde{g}_{\text{eff}}=(1+\delta)^{3/2} e^{-x \delta}, \quad \delta=m_-/M_X, \quad m_\pm=M_A \pm M_X.
\eeqn
For large $\delta$, co-annihilation is therefore exponentially suppressed and it effectively does not take place. If $\delta$ is too small and the associated co-annihilation cross-sections are large, we then have too small $\Omega_{DM}$. In the case of co-annihilation, the relative velocity is calculated from the invariant $s$ of the co-annihilating particles $X$ and $A$ as 

\bea
v=2 \sqrt{(1-m_+^2/s)(1-m_-^2/s)} \; \frac{1}{1-\frac{m_+^2 m_-^2}{s^2}} 
\eea
For $M_A=M_X \; (m_-=0)$ we recover the usual annihilation relative velocity of equation~\ref{eq:relv}. As remarked earlier, in {\tt SloopS} when calculating annihilation cross-section for DM, the phase space factor is modified such that the code returns $\sigma v$ rather than $\sigma$ which could be ill-defined for $v=0$. We give, for DM annihilation, $\sigma v$ cross-sections, in units of $\text{cm}^3 \text{s}^{-1}$. Our conversion factor to translate  $\sigma v$ expressed in $\text{GeV}^{-2} \text{m}/\text{s}$ is  
\beqn
c_0 = 2.99792 \times 3.8937966 \times 10^8 \simeq 1.16732 \times 10^9.
\eeqn


\section{$XX \to ZZ, W^+W^-$ at one-loop as a warm-up}
\label{sec:XXVVwarmup}
The  $XX \to ZZ, W^+W^-$ annihilations, with both vectors on-shell, are too efficient to account for a relic density in accordance with observation within the standard cosmological model of freeze-out. This is the reason that DM masses above $80$ GeV do not feature in Table~\ref{tab:bpfb}. However, many of the benchmark points in Table~\ref{tab:bpfb} survive because if one of the vector bosons is off-shell, the cross-sections are no longer so large. The $2 \to 3$ cross-sections $X X \to W f \bar f^\prime$ and $X X \to Z f \bar f$ do carry the salient features of the on-shell production. It is much more transparent to investigate these features on the $2\to 2$ process rather than on the more technically challenging $2\to 3$ processes that we study in~\cite{OurPaper3_2020}. Besides, $XX \to W^+W^-,ZZ$ for $M_X>80$ GeV may still be a viable model since it does not lead to overabundance but would not account for all of DM. It would then be supplemented by a new ingredient to the IDM. In any case, our aim here is to unravel some important characteristics of these $2 \to 2$ cross-sections.

Because of $SU(2)$ symmetry, $XX\to W^+W^-$ and $XX \to ZZ$ share common features. At one-loop, we concentrate first on $X X \to ZZ$ to bring up the most important features that will later help understand both our full one-loop calculation of $X X \to W f \bar f^\prime$ and $X X \to Z f \bar f$. We do this because $XX\to ZZ$ is easier to compute since there is no need to consider real corrections of the sort $XX\to W^+W^-\gamma$ that are necessary to regulate the infrared divergences of the virtual corrections. We take $M_X=100$ GeV and require all constraints of section~\ref{sec:constraints} to hold, apart from the relic density within the Planck limit. Independently of the direct detection limit we keep $\l_L$ very small, so that we are in a similar situation than in the $2\to 3$ processes we will study and whose characteristics are given by the benchmark points. For the same reason, the different choices of $M_A,\l_L$ correspond to $\l_3,\l_4,\l_5$ values similar to those found in the benchmark points for $XX\to WW^\star,Z Z^\star$. At one-loop, one needs to specify the value of $\l_2$. For each point, we take the values $\l_2=0.01,1,2$. These are the same values we considered in our study of the heavy mass IDM~\cite{Banerjee:2019luv}. The aim of this introductory analysis serves to understand the following key points:
\begin{itemize}
 
\item[\it i)] Since we are in scenarios where $M_h<2 M_X$ and can not take an OS definition of $\l_L$ based on the input $\Gamma_{h\to XX}$, we have to rely on the $\overline{MS}$ renormalisation scheme. We therefore want to investigate how large the scale dependence of the scheme is and whether we can advocate a choice for the optimum  scale.  
\item[\it ii)] Is perturbativity maintained at one-loop? Could the one-loop radiative corrections turn out to be too large for certain combinations of the underlying parameters?
\item[\it iii)] How much  do the corrections depend on $\l_2$? If the impact of $\l_2$ is not small, phenomenological considerations based on tree-level analyses should be reconsidered. 
\item[\it iv)] As an aside, we can ask whether there could be circumstances when a running of $\alpha$ (defined at $M_Z$ as assumed in {\tt micrOMEGAs}) could  account for the bulk of the corrections regardless of the value of $\l_2$. 
\item[\it v)] The scale dependence and the $\l_2$ dependence were touched upon in our study of the heavy mass IDM~\cite{Banerjee:2019luv} and we promised to investigate these issues further. We see here how the scale dependence can be derived quantitatively and how the $\l_2$ dependence is not all contained in $\bll$.
\end{itemize}

Answers to these questions will guide us when we study the $2\to 3$ processes. Not that we investigated the first three items above in our study of $\Gamma_{h\to XX}$ in the $\overline{MS}$ scheme in section~\ref{sec:hxxmsbar}. \\

To unravel some key features of the one-loop corrections, we consider 5 test points. While the DM mass, $M_X$, is fixed at $M_X=100$ GeV for all test points, we consider different values of $\l_{3,4,5}$ with the constraint that $\l_L\ll 1$. This allows us to generate different values for $\beta_{\l_L}$ as well as different values of $M_{A,H^{\pm}}$ with nonetheless $M_{A} \simeq M_{H^\pm}$ (within the $T$ parameter constraint). Characteristics of the test points are listed in Table~\ref{tab:compmul2}.
\subsection{$XX \to Z Z, W^+W^-$ at tree-level. The $\l_L$ dependence}
\begin{center}
\begin{figure}[htb]
\begin{center}
\includegraphics[width=0.48\textwidth, height=0.5\textwidth]{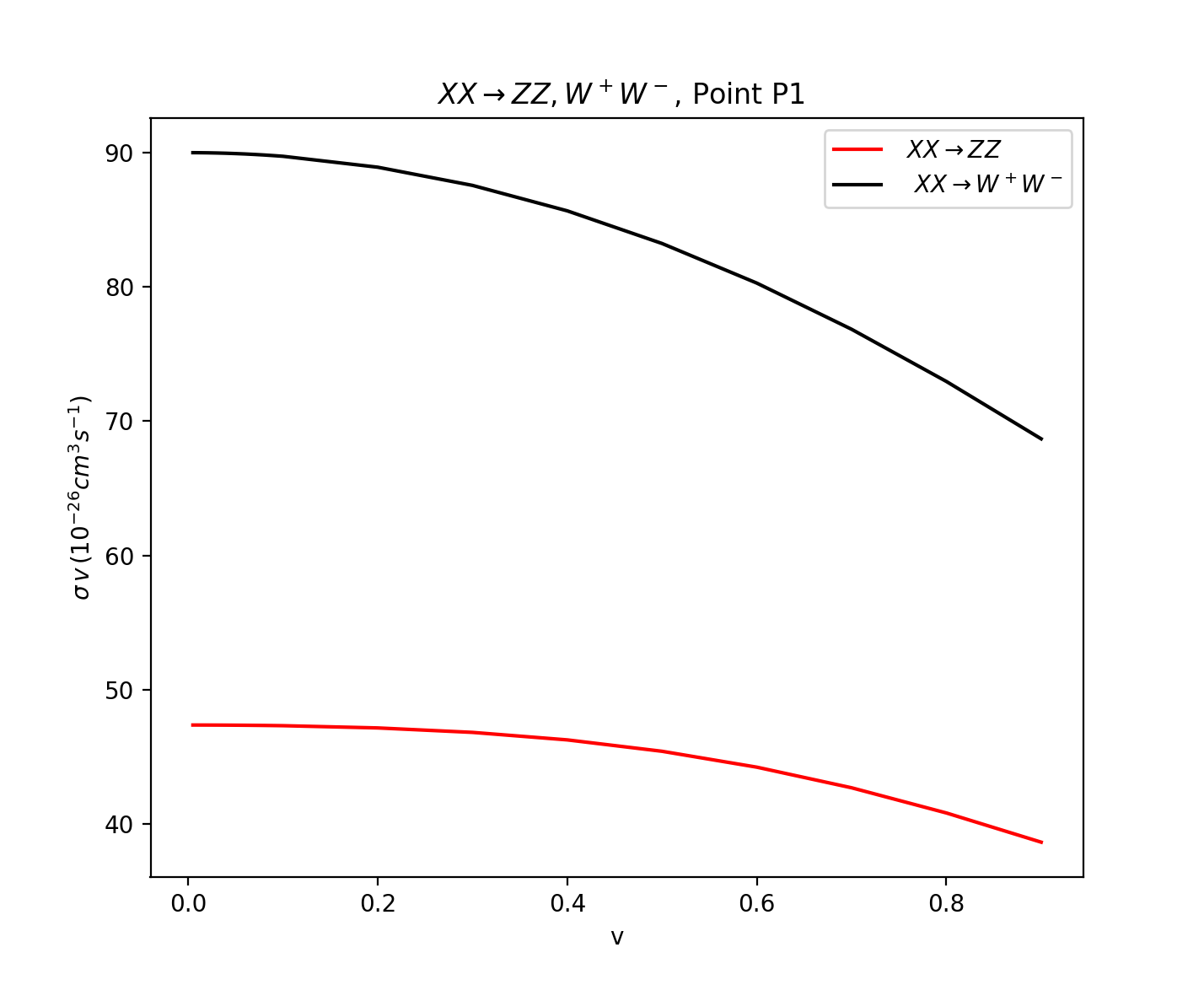}
\includegraphics[width=0.48\textwidth, height=0.5\textwidth]{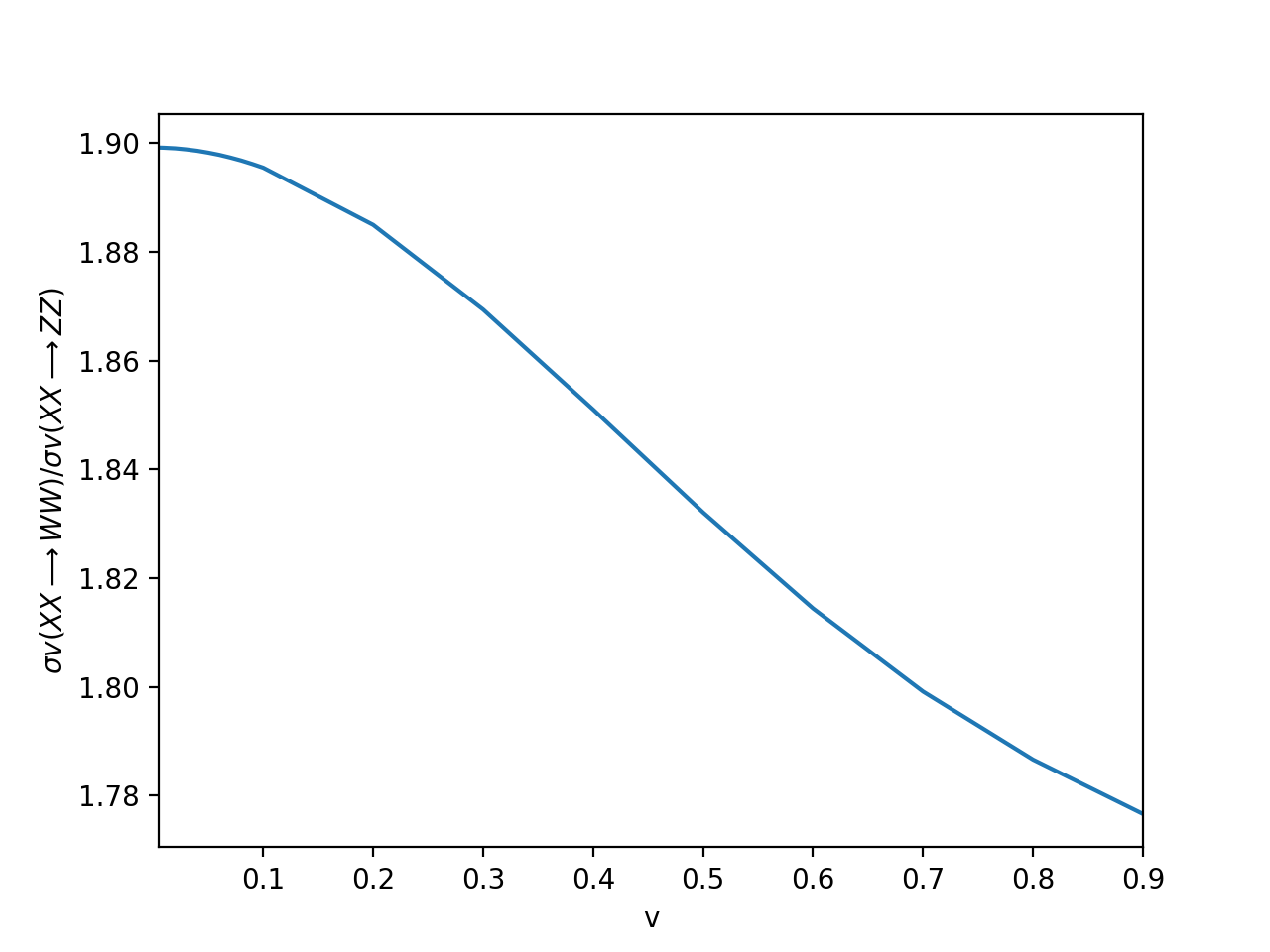}
\caption{\it The ratio $\sigma v (XX\to W^+W^-)/\sigma v (XX\to ZZ)$ as a function of the relative velocity, $v$, shown here for model P1.}
\label{fig:treesigmaXXWWoverZZ}
\end{center}
\end{figure}
\end{center}

The investigation of the tree-level cross-section gives us the $\l_L$ dependence which will then determine the scale dependence at one-loop in the $\overline{MS}$ scheme. The $\l_L$ dependence will, at one-loop, track the scale dependence which is only carried by the counterterm $\delta \l_L$.
\begin{center}
\begin{figure}[htb]
\begin{center}
\includegraphics[scale=0.4]{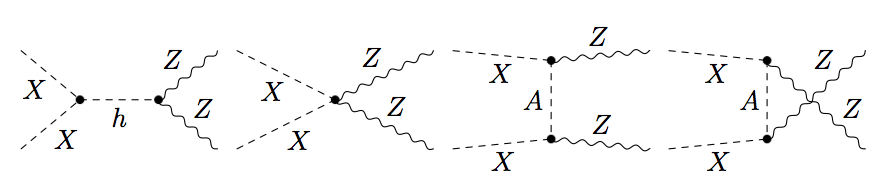}
\caption{\it Born contributions diagrams to $XX \to ZZ$. }
\label{fig:treeXXZZ}
\end{center}
\end{figure}
\end{center}
At tree-level, the $ X X \to ZZ$ Born amplitude is built up from the $s$-channel SM Higgs exchange, the quartic $XXZZ$ interaction and the exchange of $A$ in the $t$-channel, see Figure~\ref{fig:treeXXZZ}. For $X X \to W^+ W^-$, it is $H^\pm$ which is involved in the $t-$channel exchange. The tree-level cross-sections for $XX \to W^+W^-$ and $XX \to ZZ$ for test point P1 are shown in Figure~\ref{fig:treesigmaXXWWoverZZ}. Test point P1 is characterised by $\l_L=10^{-4}$ and $M_X,M_A,M_{H^\pm}=100,120,130$ GeV. The ratio between the $WW$ cross-section and the $ZZ$ cross-section is almost constant and only changes by about $6\%$ in the range $v=0-0.9$ (and less than $3\%$ in the most relevant range $v=0-0.4$), as shown in Figure~\ref{fig:treesigmaXXWWoverZZ}. For smaller $v$, $XX\to ZZ$ is slightly impacted by the threshold factor (for $M_X=100$ GeV $ZZ$ is closer to the threshold than $W^+ W^-$) which explains why the ratio is higher for small $v$ than for larger $v$s. The tree-level behaviour helps us understand why these two cross-sections share common features also at one-loop. 

As we pointed out before, the $\l_L$ dependence is not due to Higgs exchange only but it intervenes also in the $XXZZ$ quartic coupling. At the cross-section level, these contributions lead to the quadratic $\l_L^2$ dependence. The interference of these contributions with the $t$-channel amplitudes leads to a (linear) $\l_L$ dependence in the cross-section. For each point in the parameter space and for any given energy, or relative velocity, the $\l_L$ dependence of the tree-level cross-section, $\sigma_{{\rm tree}}$, can then be written as 
\beqn
\label{eq:sigmalLdep}
\sigma_{{\rm tree}}=\sigma_0 + \l_L \sigma_1 +\l_L^2 \sigma_2.
\eeqn
For $XX \to ZZ$, both Higgs exchange and the $XXZZ$ quartic coupling contribute to $\sigma_2$ whose value depends on $M_X$ and $M_h$ given a specific relative velocity. The $t$-channel $A$ exchange diagrams and part of the $XXZZ$ contribute to $\sigma_0$ and depend therefore on $M_A$ (and the SM gauge coupling). $\sigma_1$ is the result of the interference of the amplitudes contributing to $\sigma_2$ and $\sigma_0$. We choose to derive $\sigma_{0,1,2}$ numerically from {\tt SloopS}. For this purpose, we generate 3 values of $\sigma_{{\rm tree}}$  corresponding to 3 values of $\l_L$ (we take $\l_L=0,1,2$) with \underline{all other parameters fixed}. We are then able to reconstruct the $\l_L$ dependence in Equation~\ref{eq:sigmalLdep} from a fit. 
Checks of the fit are found to be excellent when comparing the result of the fit-cross-section with a direct (tree-level) calculation for a fourth value of $\l_L$ not used for the fit. Obviously, for this $2 \to 2$ process, the $\l_L$ dependence could have been derived analytically but the main reason we propose this procedure is because we follow this also for the $2\to 3$ processes~\cite{OurPaper3_2020} where an analytical expression is rather involved. \\

\subsection{$XX \to Z Z$ at one-loop: Some examples}
We start by considering $XX\to ZZ$ for a fixed value of $v$, $v=0.4$, before showing the full one-loop corrections for both $XX\to ZZ$ in the range $0<v<0.9$. At the heart of our discussion is the $\l_L$ dependence of the cross-sections which we construct according to Equation~\ref{eq:sigmalLdep}.

\begin{center}
\begin{figure}[htbp]
\begin{center}
\includegraphics[scale=0.34]{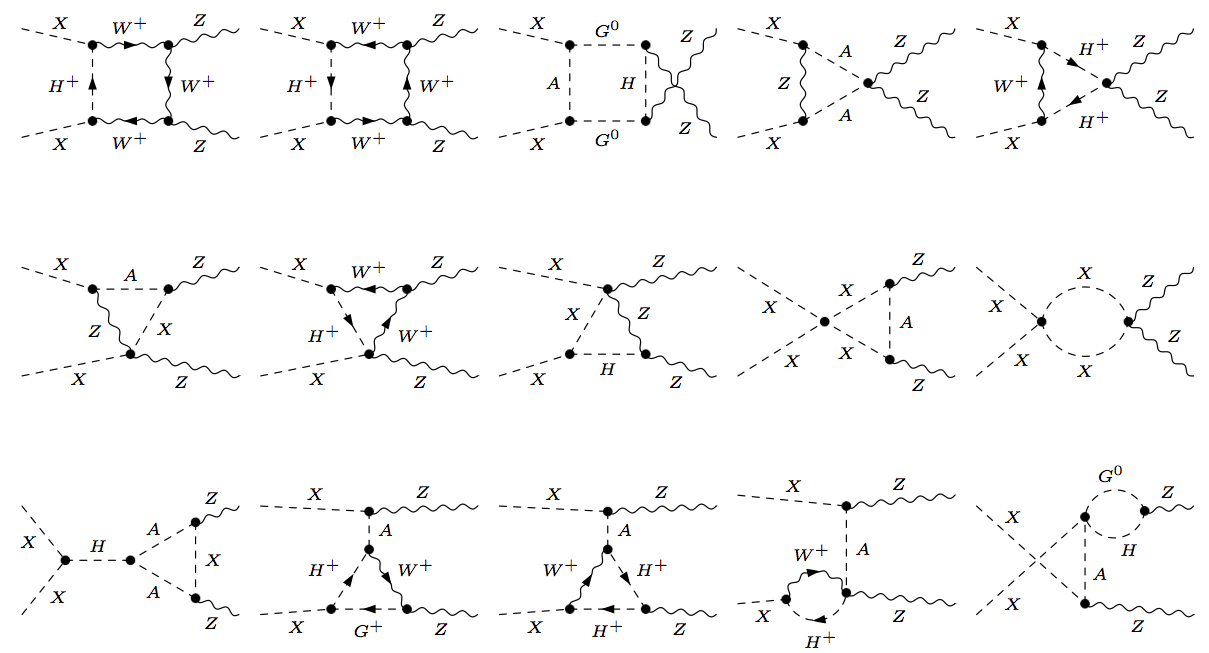}
\caption{\it A small selection of one-loop contributions diagrams to $XX \to ZZ$. We only pick up some box and triangle contributions. Note the rescattering $XX \to XX$ and $XX \to AA$ diagrams are solely with the dark sector (last 2 diagrams in the second row and the last diagram in the third row).}
\label{fig:1loopXXZZ}
\end{center}
\end{figure}
\end{center}

$XX \to ZZ$ at one-loop shows a very rich structure that accesses the parameters of the full model, in particular the $\l_2$ parameter of the dark sector. The latter shows up as rescattering effects $XX \to AA$ and $XX \to XX$, see Figure~\ref{fig:1loopXXZZ}.

Before showing the results of the full one-loop corrections in the $\overline{MS}$ scheme, let us derive the scale variation from the $\l_L$ dependence of the tree-level cross-section, $\sigma_{{\rm tree}}$ (see~\ref{eq:sigmalLdep}). The scale variation only enters through the scale variation of $\l_L$, $\partial \l_L/\partial \log \mu=-\beta_{\l_L}/16\pi^2$. Since 
\beqn
\delta \sigma_{{\rm tree}} =\frac{\partial \sigma}{\partial \l_L} \delta \l_L=(\sigma_1+2 \l_L \sigma_2) \delta \l_L,
\eeqn
one can relate the one-loop correction, $\delta \sigma$, between two scales as
\beqn
\label{eq:XXZZscaling}
\delta \sigma (\bar\mu_2)=\delta \sigma (\bar\mu_1)-\frac{1}{16 \pi^2} \Bigg(\sigma_1+2 \l_L \sigma_2\Bigg)\bll \ln (\bar \mu_2/\bar \mu_1).
\eeqn
Equation~\ref{eq:XXZZscaling} agrees perfectly with the numerical results of the full one-loop corrections that we obtain with our code. The results are presented in Table~\ref{tab:compmul2}. We could have run the code for only one value of $\mu_{{\rm dim}}$ and derived the results for any other scale through Equation~\ref{eq:XXZZscaling}. However, extracting the results for the different scales from the code is testimony that the code works very well and the implementation is consistent.\\

For later reference, with the condition that $\l_L\ll 1$, note that in terms of relative correction (normalised to the tree-level cross-section), Equation~\ref{eq:XXZZscaling} turns into 
\beqn
\label{eq:XXZZscalingRel}
\frac{\delta \sigma (\bar\mu_2)}{\sigma_{{\rm tree}}}&=&\frac{\delta \sigma (\bar\mu_1)}{\sigma_{{\rm tree}}}-\frac{1}{16 \pi^2} \Bigg(\frac{\sigma_1+2 \l_L \sigma_2}{\sigma}\Bigg)\bll \ln (\bar \mu_2/\bar \mu_1) \nonumber \\
\frac{\delta \sigma (\bar\mu_2)}{\sigma_{{\rm tree}}}&\simeq&\frac{\delta \sigma (\bar\mu_1)}{\sigma_{{\rm tree}}}-\frac{1}{16 \pi^2}\;\frac{\sigma_1}{\sigma_0} \;\bll \ln (\bar \mu_2/\bar \mu_1).
\eeqn
Equation~\ref{eq:XXZZscalingRel} indicates that the scale variation and the associated correction is large when $\bll$ is large but also when $\sigma_1/\sigma_0$ is large. For the test points P0-P5, $\sigma_1/\sigma_0$ is between $5$ and $6$.

Our results for the full one-loop corrections of the cross-section $XX \to ZZ$ for points P0-P5, evaluated at $v=0.4$,  are given in Table~\ref{tab:compmul2}. Beside listing the characteristics of the points P0-P5, Table~\ref{tab:compmul2} gives the $\l_L$ dependence of the tree-level cross-sections as well as the $\bll$ of the model such that the reader can compare the results of the analytical scale variation (Equation~\ref{eq:XXZZscalingRel}) with the numerical output of {\tt SloopS} for the full one-loop corrections for the different renormalisation scales and values of $\l_2$. In passing, note that the relative dependence of the cross-section in $\l_L$ is not large. This dependence will be stronger when we study $2 \to 3$ processes~\cite{OurPaper3_2020}. Table~\ref{tab:compmul2} is extremely instructive. It reveals many important points.

\begin{table}[h]
\centering
\tiny
\begin{tabular}{|l ||c |c| l |l|l ||}
\hline
 (Model) \hspace*{0.3cm}$\l_L$,$M_X$, $M_A$,$M_{H^\pm}$& tree  & $\bar \mu_{{\rm dim}}$&$\l_2=0.01$& $\l_2=1$ &$\l_2=2$ \\ 
& $\Delta $ &  & &   & \\ 
\hline
(P0)\hspace*{0.3cm}$\quad 0.$, $100,100,100$& 39.6681 & $M_X/2$ & 4.9264 (12.4\%)& 2.3952  (6.04\%)& -0.1616  (-0.41\%)\\
$\l_4=0., \l_5=0., \l_3=0.$& 5.1608 (13\%) & $M_X=M_A$ & 4.4240 (11.15\%)& 1.8928(4.77\%) & -0.6639 (-1.67\%) \\
$\bll= 0.469$& & $ 2 M_X$ & 3.9217 (9.89\%) & 1.3905(3.51\%) & -1.1663 (-2.94\%)\\
$\sigma_{{\rm tree}}=39.6681 + 244.184 \l_L + 435.584 \l_L^2$&&&&& \\ 
$\;\;\;\;\;\;\;\; \simeq 39.67 (1+6.2\l_L+11\l_L^2)  $ &&&&& \\\hline
\hline
(P1)\hspace*{0.3cm}$\quad 10^{-4}$, $100,120,130$& 46.2668&    $M_X/2$ & 5.9161 (12.78\%)& 4.6733 (10.10\%)& 3.4179 (7.39\%)\\
$\l_4=-0.150, \l_5=-0.070, \l_3=0.221$&6.0173 (13\%)&  $M_X$ & 5.0211(10.85\%) & 2.3894 (5.16\%)& -0.2689 (-0.58\%)\\
$\bll= 0.731+1.165\l_2$ & &$M_A$& 4.7862 (10.34\%)& 1.7887 (3.87\%)& -1.239 (-2.68\%)\\
$\sigma_{{\rm tree}}=46.2394 + 274.344 \l_L+ 435.584 \l_L^2$ &&$2 M_X$ & 4.1261 (8.91\%)& 0.1055 (0.23\%)& -3.9557 (-8.55\%)\\ 
$\;\;\;\;\;\;\;\; \simeq 46.24 (1+5.9\l_L+9.4\l_L^2)  $ &&&&& \\ 
\hline
(P2) \hspace*{0.3cm}  $5\times10^{-4}$, $100,200,200$&62.0202 & $M_X/2$& 16.2987 (26.28\%)& 26.7585 (43.14\%) & 37.3240 (60.18\%)\\
$\l_4=\l_5=-0.479, , \l_3=0.959$& 8.0576 (13\%)& $M_X$&10.5437 (17\%)& 12.8096(10.65\%) & 15.0984(24.34\%) \\
$\bll= 3.947+5.759\l_2$& & $2 M_X=M_A$& 4.7886 (7.72\%)& -1.1393 (-1.84\%)& -7.1271 (-11.49\%)\\
$\sigma_{{\rm tree}}=61.8565 + 326.957 \l_L + 435.584 \l_L^2$&&&&& \\
$\;\;\;\;\;\;\;\; \simeq 61.86 (1+5.3\l_L+7\l_L^2)  $ &&&&& \\ \hline
\hline
(P3) \hspace*{0.3cm}   $5\times10^{-4}$, $100,250,250$& 66.4210& $M_X/2$& 39.3295 (59.21\%)&\bf { 66.6048 }& {\bf 94.1555}\\
$\l_4=\l_5=-0.839, \l_3=1.678$& 8.6297 (13\%)& $M_X$& 24.0102 (36.15\%) & 36.4168 (59.34\%)& 48.9487(73.69\%) \\ 
$\bll= 10.175+10.075\l_2$& & $2 M_X$& 8.6914 (13.08\%)& 6.2276 (9.38\%) & 3.7388 (5.63\%)\\
$\sigma_{{\rm tree}}=66.25 + 339.199 \l_L + 435.583 \l_L^2$  & & $M_A$& 3.7592 (5.66\%)& -3.4896 (-5.25\%)& -10.8115 (-16.28\%) \\
$\;\;\;\;\;\;\;\; \simeq 66.25 (1+5.2\l_L+6.6\l_L^2)  $ &&&&& \\ \hline
(P4) \hspace*{0.3cm} $ 10^{-4}$, $100,500,500$&73.0022 &  $M_X/2$& {\bf 944.4573} & {\bf 1170.1050 }& {\bf 1398.0319}\\
$\l_4=\l_5=-3.836, \l_3=7.671$& 9.4949 (13\%)&$M_X$&{\bf 657.6667}&{\bf 811.9868} &{\bf 967.8657}\\
$\bll= 182.758+46.029\l_2$&&$2M_X$&{\bf 370.8766}&{\bf 453.8675}&{\bf 537.6966}\\
$\sigma_{{\rm tree}}=72.9665 + 356.520 \l_L + 435.584 \l_L^2$ &&$M_A$&-8.2405 (-11.28\%)&-19.5371 (-26.76\%)&-30.9477 (-42.38\%)\\
$\;\;\;\;\;\;\;\; \simeq 72.97 (1+4.9\l_L+6\l_L^2)  $ &&&&& \\\hline \hline
\end{tabular}
\normalsize
\caption{\label{tab:compmul2}{\it Tree-level cross-section and one-loop correction, $\delta \sigma(\bar \mu_{{\rm dim}},\l_2)$ for different values of $\l_2$ and for different renormalisation scales, $\bar \mu_{{\rm dim}}$, for $\l_L$ for $XX \to ZZ$ and $v=0.4$ ($\sqrt{s}=204.124$). The percentage change of the one-loop correction is also given in parenthesis unless the value is higher than 100\% in which case the result is highlighted in bold as a warning for a breakdown of one-loop perturbativity. The loop results (from the numerical evaluation of the automated calculations) are given with sufficient accuracy in order to compare the results against those of the analytical computation given in Equation~\ref{eq:XXZZscaling}. The first entry in the second column is the tree-level cross-section while the second entry in the second column corresponds to $\Delta=\sigma_{{\rm tree}}(\alpha^{-1}(M_Z^2)=128.907)-\sigma_{{\rm tree}}(\alpha^{-1}(0)=137.036)$, the ``correction" that corresponds to using $\alpha(M_Z^2)$ instead of $\alpha$ in the Thomson limit (tree-level). The value of $\Delta$ needs to be compared to the result of the full one-loop calculation. The first column gives the parameters of the model, with $M_X=100$ GeV for all points, and the corresponding $\l_{3,4,5}$. In the same column, we also give $\beta_{\l_L}$ and a fit of the tree-level cross-section as a function of $\l_L$ (see text). Cross-sections are for $\sigma v$ in units of $10^{-26}$cm$^3\;s^{-1}$}. 
}
\end{table}

\begin{itemize}
\item The rescattering effects within the dark sector, through the $\l_2$ dependence, are confirmed. The one-loop calculation shows that $\l_2$ enters the prediction of the annihilation cross-section not only through the running of the coupling $\l_L$, through $\bll$, but there is also an added genuine one-loop contribution. Indeed, we have chosen test point P0 because its $\bll$ does not depend on $\l_2$. Therefore, the results for P0 do show a dependence of the correction on $\l_2$ that is not an effect of the scale dependence of $\l_L$.  For P0, this dependence can be derived easily from Table~\ref{tab:compmul2}, by taking the difference of any two columns (that is the difference between two values of $\l_2$ evaluated at the same scale). We find that $\delta \sigma (\l_2)=-2.56 \; \l_2$ or in terms of relative correction $-6.5\; \l_2 (\%)$, which is not negligible for values of $\l_2>1$. For the other test points, there is also a $\l_2$ dependence that enters through $\bll$. Therefore, only a full one-loop correction captures the full $\l_2$ dependence, beyond the dependence contained in $\bll$. The $\l_2$ dependence is important. 
\item The running of $\alpha$ at the scale of the process, of order $M_Z$, does not account for the full one-loop correction. It can be considered as an acceptable description only for an almost vanishing $\l_2$ {\underline {and} very small} $\l_{3,4,5}$. In fact, for test point P0, with $\l_2=2$, the corrections practically vanish showing that extra corrections totally offset the correction from the running of $\alpha$. Therefore, the use of a running $\alpha$ is of very limited applicability. 
\item As the split between the DM mass and the mass of the other scalars increases, the one-loop correction and the scale dependence increase. This can be (mostly) understood on the basis of the corresponding value of $\bll$. A large $\bll$ is a harbinger of a large scale uncertainty. An inadequate choice of the scale can further exacerbate a large correction driven by $\bll$. $\bll$ increases with $\l_2$ but even for $\l_2 \sim 0$, $\bll$ can be large leading to a large correction and casting doubt on the perturbative expansion. For example, P3 and most particularly P4 have very large scale uncertainty for all values of $\l_2$. Based on these one-loop analyses, we make the proposal that such models fail the perturbativity test and should therefore not be considered even though they may pass the tree-level based experimental limits. One should not consider models with $\bll$ larger than 20. For a starter, Table~~\ref{tab:compmul2} shows that when one reaches higher values of $\bll$, the {\it corrections} reach more than 100\% unless one carefully picks up an optimal scale. In any case, for large values of $\bll$, one is subject to violent variations even if one moves slightly away from the optimum scale. Second, if one takes QCD as example with $g_s^2= 4\pi \alpha_s$, one has (in our notation), $\beta_{g_s} \sim -13$ while (for the most relevant quantity), $\beta_{\alpha_s}=-2.5$ at energy scales where $\alpha_s$ is perturbative ($\alpha_s \sim 0.12$). For $Q^2\sim (1{\rm GeV})^2$ when QCD is non-perturbative, $\beta_{\alpha_s} \sim 20$. Therefore, $\bll \sim 20$ should be a good indicator for the onset of the non-perturbative regime. This indicator when translated in terms of the underlying parameters of the IDM suggests that we should not consider IDM benchmarks points to be theoretically valid if $M_A (M_{H^\pm}) > M_X +150 \; \text{GeV}$, otherwise the corresponding $\l_{3,4,5}$ correspond to values that lead to $\bll$ in excess of $20$. This requirement of one-loop perturbativity reduces the IDM parameter space considerably. We take this conclusion into account to reduce the possibilities for the  benchmark points of Table~\ref{tab:bpfb}. We consider the benchmark points B, C, D, H to be  theoretically unreliable.
\item In our study of the scale dependence for the partial width, $\Gamma_{h\to XX}$, in section~\ref{sec:renormalisation}, we advocated the use of $\mu={\rm max}(2 M_X,M_A)$ as the optimum scale where the radiative corrections were minimised. Table~\ref{tab:compmul2} suggests a very similar behaviour. For the annihilations processes, the typical energy scale of the problem set by kinematics is $\sqrt{s} \simeq 2 M_X$ while the the typical internal mass is $M_A$. Although we are dealing with a process involving 4-point functions, the arguments put forward in the case of $h\to XX$ remain the same especially because the scale dependence is embedded in 2- and 3-point functions. We find that $\mu={\rm max}(2 M_X,M_A)$ is once again the best scale. Scales much smaller or larger than this choice lead to very large corrections. This is most flagrant for test points P3 and P4 that we dismiss on the ground of perturbativity, to the point that following this prescription would lead to acceptable corrections if we choose $\mu=M_A$. For the ``perturbative" test points P0, P1, and P2, the prescription is most evident for P2. The indication from P1 with $\l_2=2$ is misleading although the corrections are small for all scales ($\bll$ is small here), $\mu=2M_X$ does not give the smallest correction because of the $\l_2$ contribution not originating from $\bll$. For $\l_2=0.01$, $\mu=2M_X$ gives the smallest correction. These observations will be confirmed also when we study $XX \to Z f \bar f$ and $XX \to W f \bar f^\prime$~\cite{OurPaper3_2020}. 
\end{itemize}

We also find that the $\l_L$ dependence of the $XX \to W^+ W^-$ is practically the same as the $XX \to ZZ$ cross-section. $XX \to W^+ W^-$ will therefore exhibit the same scale dependence as we will see next. 
 
\subsection{$XX \to ZZ$ and $XX \to W^+W^-$ at one-loop as functions of the relative velocity}
The particular study at $v=0.4$ summarises, in fact, the analyses for the whole range of $v$. We therefore briefly discuss here the results for point 1 for the annihilations into both $ZZ$ and $WW$.
\begin{center}
\begin{figure}[tbh]
\includegraphics[width=0.33\textwidth, height=0.4\textwidth]{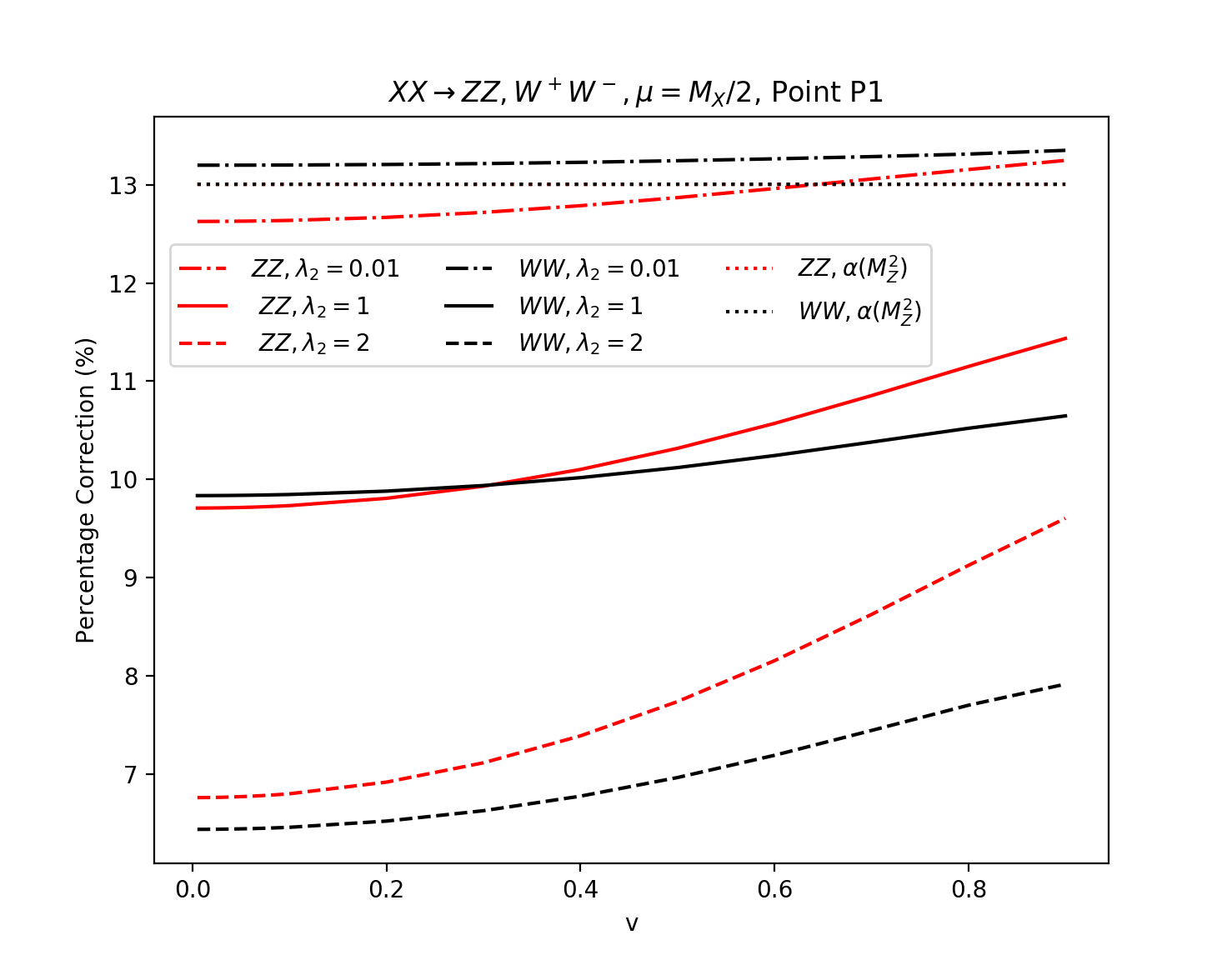}
\includegraphics[width=0.33\textwidth, height=0.4\textwidth]{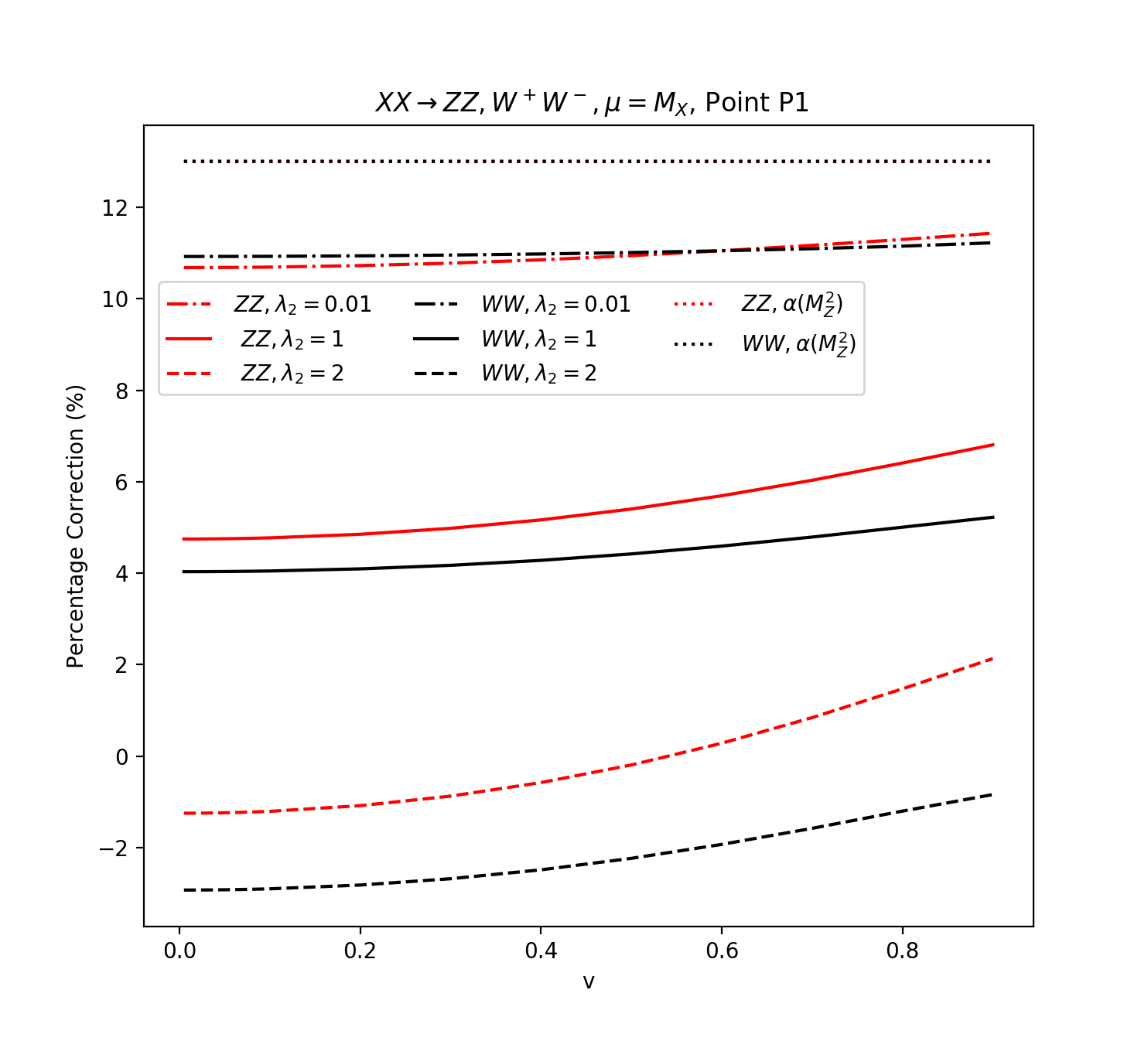}
\includegraphics[width=0.33\textwidth, height=0.4\textwidth]{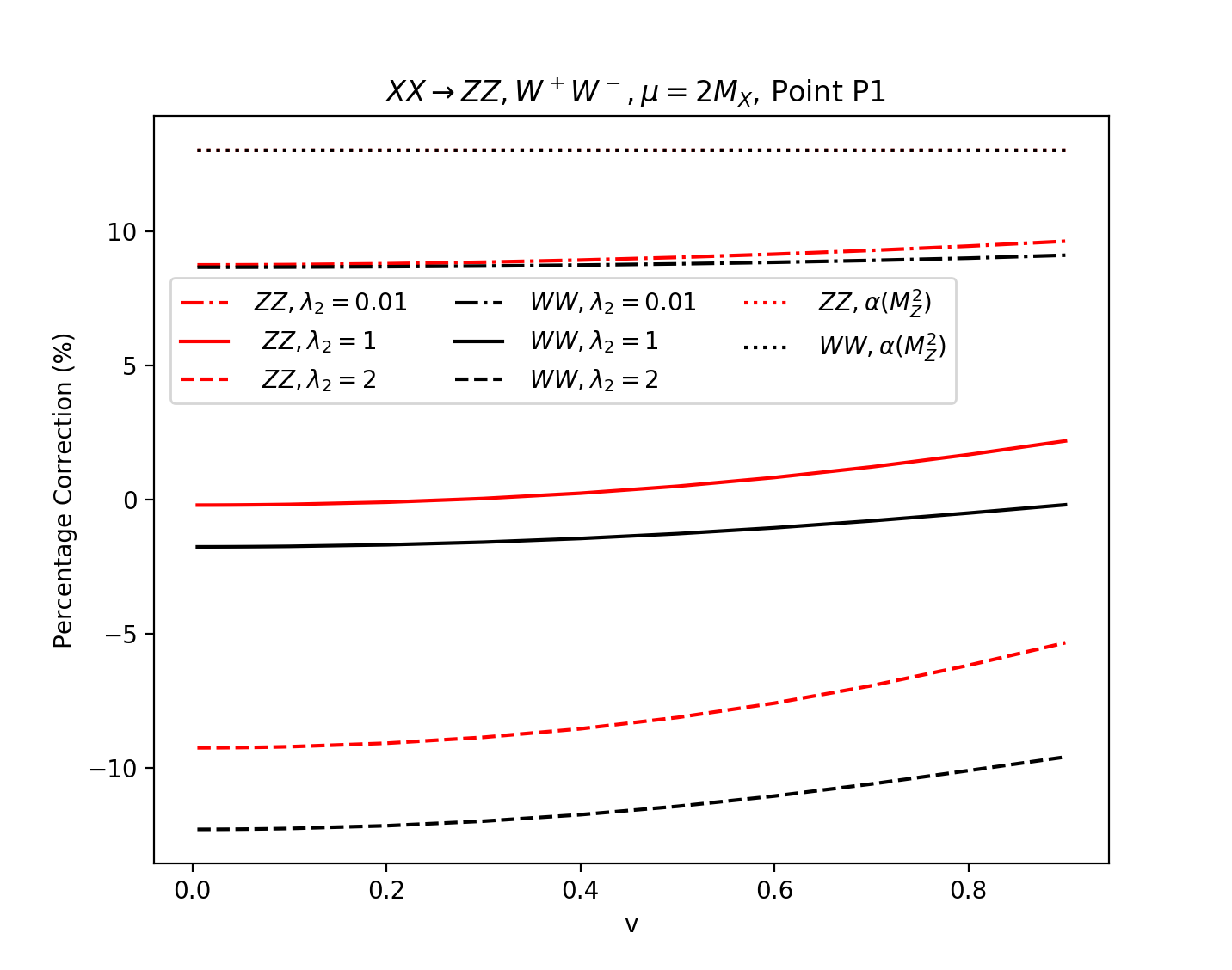}
\caption{\label{fig:XXVV1loopRel}\it The relative (to the tree-level) full one-loop corrections to the cross-sections $XX \to W^+W^-$ and $ZZ$ as functions of the relative velocity for test point P1 with $\mu=M_X/2,M_X,2 M_X$ (from left to right). Tree values of $\l_2$ are considered. The percentage deviation due to choosing as input $\alpha(M_Z^2)$ instead of $\alpha(0)$ is also shown. This correction ($\sim 13\%$) is the same for both $ZZ$ and $WW$. It is therefore superimposed for the two cross-sections and appears as a common line.}
\end{figure}
\end{center}
Recall that P1 has a small $\bll$ and therefore the scale variation is modest not only for $v=0.4$, as we discussed, but is expected to hold for other values of $v$. Indeed, Figures~\ref{fig:XXVV1loopRel} show that the radiative corrections, for the range of $v$ relevant for the calculation of the relic density, are within a couple of per-cent of the results obtained with $v=0.4$, for all cases of the scale $\mu$ and the parameter $\l_2$. The radiative corrections increase slightly, within a $2\%$ margin, with the relative velocity. The increase affects more $ZZ$ than $W^+W^-$ as $v$ increases and is due to the fact that $ZZ$ is more sensitive to the threshold for $ZZ$ production for $M_X=100$ GeV. Otherwise, observe that, as expected in terms of relative corrections, the results are practically the same for both $ZZ$ and $W^+W^-$ production when the same configuration of the scale and $\l_2$ is taken. More importantly, what the full one-loop correction teaches us is that the "improved" tree-level cross-section, with the use of $\alpha(M_Z^2)$, as assumed by default in {\tt micrOMEGAs}, is a good approximation only for vanishingly small $\l_2$. This "improved" cross-section can not, by construction, catch the genuine one-loop effects from the full $\l_2$ dependence. The prediction of the $\alpha(M_Z^2)$ approximation can be quite far from the full one-loop correction for large values of $\l_2$. For instance, for $\mu=2M_X$ and $\l_2=2$, the $\alpha(M_Z^2)$ cross-sections are practically $20\%$ higher than the full one-loop correction. In general, we see that measured from the tree-level cross-section, the corrections decrease as $\l_2$ increases and they also decrease as the scale increases, a behaviour which is fully explained by our $\l_L, \bll$ discussion exposed through Equation~\ref{eq:XXZZscalingRel}. Therefore, we recommend for such scenarios, with small $\bll$, to apply a theoretical uncertainty to the prediction of {\tt micrOMEGAs} ranging from $-20\%$ to $-4\%$ to account for the scale uncertainty and the $\l_2$ dependence that a tree-level calculation is not sensitive to. 


\section{Conclusions}
\label{sec:conclusions}
Compared to a much studied model like the MSSM (minimal supersymmetric model), the IDM with the addition of one scalar doublet to the SM is quite simple and economical in terms of the number of parameters it involves. Yet, it has a rich structure and constitutes an excellent laboratory to address many technical issues about renormalisation of a BSM model. Namely, what physical input parameters one could define to carry a full one-loop calculation. Using the masses of the new fields (particles) of the model seems a most unambiguous scheme to define the model. However, even if all masses are given, we still need two more parameters, $\l_L$ (or some other combination of the underlying parameters of the Lagrangian) and $\l_2$, to compute observables, most importantly cross-sections. $\l_L$ which measures the strength of the coupling of the SM Higgs to the DM particle, is needed for the (tree-level) calculation of the annihilation of DM to SM particles, besides the gauge couplings that we define from SM observables. The quartic coupling, $\l_2$, represents interactions solely within the dark sector. Nonetheless, loop effects introduce an important $\l_2$ dependence of the annihilation cross-sections. This indirect one-loop effect introduces therefore an uncertainty in the prediction of an observable such as the relic density. $\l_L$ also introduces an uncertainty of a different sort. Either this coupling is extracted from the partial invisible width of the Higgs, phase space allowing, or as we advocated in this paper an $\overline{{\rm MS}}$ scheme is prescribed to allow a full one-loop calculation. The latter introduces a renormalisation scale dependence in the one-loop predictions. If the $\overline{{\rm MS}}$ prescription is applied to $\l_L$ only, we have shown how the scale dependence can be tracked through the $\beta$ constant for $\l_L$ ($\bll$) and the $\l_L$ parametrisation of the tree-level observable. We have also argued how to reduce the scale uncertainty by choosing an optimal scale which we conjecture to be close to the largest scale involved in the process. The overall theoretical uncertainty should be estimated by varying the scale around this optimal scale and also by varying $\l_2$, a quantity the annihilation of DM to SM particles does not depend on. The study of the theoretical uncertainty at one-loop has also led us to introduce a new criterion for the perturbativity requirement: only configurations with small enough $\bll$ qualify.
From the point of view of the relic density calculation, the IDM, even in the narrow range of low masses ($55< M_X< 75$ GeV), involves the main known mechanisms in the freeze-out scenario: annihilation in the continuum, co-annihilation, annihilation through a resonance. We have conducted a very thorough investigation on the allowed parameter space of the model which includes new LHC and direct detection data to delimit the range of the low mass IDM scenarios and the DM annihilation mechanisms that are involved. A set of representative IDM points covering these mechanisms was presented. It will serve as a starting point to conduct full one-loop calculations in these three scenarios. $XX \to W f \bar f^\prime $ (a $2 \to 3$ process) at one-loop which has never been attended before, will figure in all three scenarios but with varying degree of importance. Salient features of this process are contained in the $2\to 2$ processes $XX \to W^+ W^-$ and $ZZ$ which we have studied in this paper. Crucial technical process specific issues, beyond the general renormalisation procedure presented at some length here, are presented in the accompanying papers that cover each of the three scenarios. The calculation of the relic density at one-loop, when the mechanism is dominated by co-annihilation, will be presented in Ref.~\cite{OurPaper2_2020}. If we set aside the induced sub-dominant production through $XX \to W f \bar f^\prime $ (a $2 \to 3$ process), the dominant mechanism is $AX$ co-annihilation to a fermion pair. In Ref.~\cite{OurPaper3_2020}, we consider 3 benchmark points to present results for $XX \to W f \bar f^\prime, XX \to Z f \bar f$ and how the corrected cross-sections translate into the calculation of the relic density. Finally in Ref.~\cite{OurPaper4_2020}, the resonance region, we  will show  how to extend the OS scheme we detailed in the present paper by supplementing a complex scheme that avoids the issue of double counting in the presence of a width. The width, necessary in the tree-level calculation, is in fact induced at the loop level calculation since the width represents the imaginary part of the self-energy contribution. Again, the IDM can be a good example which illustrates how a loop calculation for a process that proceeds through a resonance, should be conducted.


\acknowledgments
We are grateful to Guillaume Chalons who contributed to the early stages of this work. We thank Jack Y. Araz  for assistance regarding aspects of {\tt MadAnalysis 5}. HS is supported by the National Natural Science Foundation of China (Grant No.12075043,  No. 11675033). He warmly thanks the CPTGA and LAPTh for support during his visit to France in 2019 when this work was initiated. SB is grateful for the support received from IPPP, Durham, UK, where most of this work was performed. SB also thanks LAPTh for the support provided during his visit, when this work started. NC is financially supported by IISc (Indian Institute of Science, Bangalore, India) through the C.V.~Raman postdoctoral fellowship. NC also acknowledges support from DST, India, under grant number IFA19-PH237 (INSPIRE Faculty Award).

\bibliography{../../NLO_IDM570}

\providecommand{\href}[2]{#2}\begingroup\raggedright\begin{thebibliography}{100}

\bibitem{Ade:2015xua}
{\scshape Planck} collaboration, P.~A.~R. Ade et~al., \emph{{Planck 2015
  results. XIII. Cosmological parameters}},
  \href{http://dx.doi.org/10.1051/0004-6361/201525830}{\emph{Astron.
  Astrophys.} {\bf 594} (2016) A13},
  [\href{http://arxiv.org/abs/1502.01589}{{\tt 1502.01589}}].

\bibitem{Deshpande:1977rw}
N.~G. Deshpande and E.~Ma, \emph{{Pattern of Symmetry Breaking with Two Higgs
  Doublets}}, \href{http://dx.doi.org/10.1103/PhysRevD.18.2574}{\emph{Phys.
  Rev.} {\bf D18} (1978) 2574}.

\bibitem{Barbieri:2006dq}
R.~Barbieri, L.~J. Hall and V.~S. Rychkov, \emph{{Improved naturalness with a
  heavy Higgs: An Alternative road to LHC physics}},
  \href{http://dx.doi.org/10.1103/PhysRevD.74.015007}{\emph{Phys. Rev.} {\bf
  D74} (2006) 015007}, [\href{http://arxiv.org/abs/hep-ph/0603188}{{\tt
  hep-ph/0603188}}].

\bibitem{Hambye:2007vf}
T.~Hambye and M.~H.~G. Tytgat, \emph{{Electroweak symmetry breaking induced by
  dark matter}},
  \href{http://dx.doi.org/10.1016/j.physletb.2007.11.069}{\emph{Phys. Lett.}
  {\bf B659} (2008) 651--655}, [\href{http://arxiv.org/abs/0707.0633}{{\tt
  0707.0633}}].

\bibitem{Banerjee:2019luv}
S.~Banerjee, F.~Boudjema, N.~Chakrabarty, G.~Chalons and H.~Sun, \emph{{Relic
  density of dark matter in the inert doublet model beyond leading order: The
  heavy mass case}},
  \href{http://dx.doi.org/10.1103/PhysRevD.100.095024}{\emph{Phys. Rev.} {\bf
  D100} (2019) 095024}, [\href{http://arxiv.org/abs/1906.11269}{{\tt
  1906.11269}}].

\bibitem{Hisano:2002fk}
J.~Hisano, S.~Matsumoto and M.~M. Nojiri, \emph{{Unitarity and higher order
  corrections in neutralino dark matter annihilation into two photons}},
  \href{http://dx.doi.org/10.1103/PhysRevD.67.075014}{\emph{Phys. Rev.} {\bf
  D67} (2003) 075014}, [\href{http://arxiv.org/abs/hep-ph/0212022}{{\tt
  hep-ph/0212022}}].

\bibitem{Hisano:2003ec}
J.~Hisano, S.~Matsumoto and M.~M. Nojiri, \emph{{Explosive dark matter
  annihilation}},
  \href{http://dx.doi.org/10.1103/PhysRevLett.92.031303}{\emph{Phys. Rev.
  Lett.} {\bf 92} (2004) 031303},
  [\href{http://arxiv.org/abs/hep-ph/0307216}{{\tt hep-ph/0307216}}].

\bibitem{Hisano:2004ds}
J.~Hisano, S.~Matsumoto, M.~M. Nojiri and O.~Saito, \emph{{Non-perturbative
  effect on dark matter annihilation and gamma ray signature from galactic
  center}}, \href{http://dx.doi.org/10.1103/PhysRevD.71.063528}{\emph{Phys.
  Rev.} {\bf D71} (2005) 063528},
  [\href{http://arxiv.org/abs/hep-ph/0412403}{{\tt hep-ph/0412403}}].

\bibitem{Hambye:2009pw}
T.~Hambye, F.~S. Ling, L.~Lopez~Honorez and J.~Rocher, \emph{{Scalar Multiplet
  Dark Matter}}, \href{http://dx.doi.org/10.1007/JHEP05(2010)066,
  10.1088/1126-6708/2009/07/090}{\emph{JHEP} {\bf 07} (2009) 090},
  [\href{http://arxiv.org/abs/0903.4010}{{\tt 0903.4010}}].

\bibitem{Biondini:2017ufr}
S.~Biondini and M.~Laine, \emph{{Re-derived overclosure bound for the inert
  doublet model}}, \href{http://dx.doi.org/10.1007/JHEP08(2017)047}{\emph{JHEP}
  {\bf 08} (2017) 047}, [\href{http://arxiv.org/abs/1706.01894}{{\tt
  1706.01894}}].

\bibitem{LopezHonorez:2006gr}
L.~Lopez~Honorez, E.~Nezri, J.~F. Oliver and M.~H.~G. Tytgat, \emph{{The Inert
  Doublet Model: An Archetype for Dark Matter}},
  \href{http://dx.doi.org/10.1088/1475-7516/2007/02/028}{\emph{JCAP} {\bf 0702}
  (2007) 028}, [\href{http://arxiv.org/abs/hep-ph/0612275}{{\tt
  hep-ph/0612275}}].

\bibitem{Cao:2007rm}
Q.-H. Cao, E.~Ma and G.~Rajasekaran, \emph{{Observing the Dark Scalar Doublet
  and its Impact on the Standard-Model Higgs Boson at Colliders}},
  \href{http://dx.doi.org/10.1103/PhysRevD.76.095011}{\emph{Phys. Rev.} {\bf
  D76} (2007) 095011}, [\href{http://arxiv.org/abs/0708.2939}{{\tt
  0708.2939}}].

\bibitem{Agrawal:2008xz}
P.~Agrawal, E.~M. Dolle and C.~A. Krenke, \emph{{Signals of Inert Doublet Dark
  Matter in Neutrino Telescopes}},
  \href{http://dx.doi.org/10.1103/PhysRevD.79.015015}{\emph{Phys. Rev.} {\bf
  D79} (2009) 015015}, [\href{http://arxiv.org/abs/0811.1798}{{\tt
  0811.1798}}].

\bibitem{Lundstrom:2008ai}
E.~Lundstrom, M.~Gustafsson and J.~Edsjo, \emph{{The Inert Doublet Model and
  LEP II Limits}},
  \href{http://dx.doi.org/10.1103/PhysRevD.79.035013}{\emph{Phys. Rev.} {\bf
  D79} (2009) 035013}, [\href{http://arxiv.org/abs/0810.3924}{{\tt
  0810.3924}}].

\bibitem{Andreas:2009hj}
S.~Andreas, M.~H.~G. Tytgat and Q.~Swillens, \emph{{Neutrinos from Inert
  Doublet Dark Matter}},
  \href{http://dx.doi.org/10.1088/1475-7516/2009/04/004}{\emph{JCAP} {\bf 0904}
  (2009) 004}, [\href{http://arxiv.org/abs/0901.1750}{{\tt 0901.1750}}].

\bibitem{Arina:2009um}
C.~Arina, F.-S. Ling and M.~H.~G. Tytgat, \emph{{IDM and iDM or The Inert
  Doublet Model and Inelastic Dark Matter}},
  \href{http://dx.doi.org/10.1088/1475-7516/2009/10/018}{\emph{JCAP} {\bf 0910}
  (2009) 018}, [\href{http://arxiv.org/abs/0907.0430}{{\tt 0907.0430}}].

\bibitem{Dolle:2009ft}
E.~Dolle, X.~Miao, S.~Su and B.~Thomas, \emph{{Dilepton Signals in the Inert
  Doublet Model}},
  \href{http://dx.doi.org/10.1103/PhysRevD.81.035003}{\emph{Phys. Rev.} {\bf
  D81} (2010) 035003}, [\href{http://arxiv.org/abs/0909.3094}{{\tt
  0909.3094}}].

\bibitem{Nezri:2009jd}
E.~Nezri, M.~H.~G. Tytgat and G.~Vertongen, \emph{{e+ and anti-p from inert
  doublet model dark matter}},
  \href{http://dx.doi.org/10.1088/1475-7516/2009/04/014}{\emph{JCAP} {\bf 0904}
  (2009) 014}, [\href{http://arxiv.org/abs/0901.2556}{{\tt 0901.2556}}].

\bibitem{Miao:2010rg}
X.~Miao, S.~Su and B.~Thomas, \emph{{Trilepton Signals in the Inert Doublet
  Model}}, \href{http://dx.doi.org/10.1103/PhysRevD.82.035009}{\emph{Phys.
  Rev.} {\bf D82} (2010) 035009}, [\href{http://arxiv.org/abs/1005.0090}{{\tt
  1005.0090}}].

\bibitem{Gong:2012ri}
J.-O. Gong, H.~M. Lee and S.~K. Kang, \emph{{Inflation and dark matter in two
  Higgs doublet models}},
  \href{http://dx.doi.org/10.1007/JHEP04(2012)128}{\emph{JHEP} {\bf 04} (2012)
  128}, [\href{http://arxiv.org/abs/1202.0288}{{\tt 1202.0288}}].

\bibitem{Gustafsson:2012aj}
M.~Gustafsson, S.~Rydbeck, L.~Lopez-Honorez and E.~Lundstrom, \emph{{Status of
  the Inert Doublet Model and the Role of multileptons at the LHC}},
  \href{http://dx.doi.org/10.1103/PhysRevD.86.075019}{\emph{Phys. Rev.} {\bf
  D86} (2012) 075019}, [\href{http://arxiv.org/abs/1206.6316}{{\tt
  1206.6316}}].

\bibitem{Swiezewska:2012eh}
B.~Swiezewska and M.~Krawczyk, \emph{{Diphoton rate in the inert doublet model
  with a 125 GeV Higgs boson}},
  \href{http://dx.doi.org/10.1103/PhysRevD.88.035019}{\emph{Phys. Rev.} {\bf
  D88} (2013) 035019}, [\href{http://arxiv.org/abs/1212.4100}{{\tt
  1212.4100}}].

\bibitem{Wang:2012zv}
L.~Wang and X.-F. Han, \emph{{LHC diphoton Higgs signal and top quark
  forward-backward asymmetry in quasi-inert Higgs doublet model}},
  \href{http://dx.doi.org/10.1007/JHEP05(2012)088}{\emph{JHEP} {\bf 05} (2012)
  088}, [\href{http://arxiv.org/abs/1203.4477}{{\tt 1203.4477}}].

\bibitem{Goudelis:2013uca}
A.~Goudelis, B.~Herrmann and O.~Stal, \emph{{Dark matter in the Inert Doublet
  Model after the discovery of a Higgs-like boson at the LHC}},
  \href{http://dx.doi.org/10.1007/JHEP09(2013)106}{\emph{JHEP} {\bf 09} (2013)
  106}, [\href{http://arxiv.org/abs/1303.3010}{{\tt 1303.3010}}].

\bibitem{Arhrib:2013ela}
A.~Arhrib, Y.-L.~S. Tsai, Q.~Yuan and T.-C. Yuan, \emph{{An Updated Analysis of
  Inert Higgs Doublet Model in light of the Recent Results from LUX, PLANCK,
  AMS-02 and LHC}},
  \href{http://dx.doi.org/10.1088/1475-7516/2014/06/030}{\emph{JCAP} {\bf 1406}
  (2014) 030}, [\href{http://arxiv.org/abs/1310.0358}{{\tt 1310.0358}}].

\bibitem{Krawczyk:2013jta}
M.~Krawczyk, D.~Sokolowska, P.~Swaczyna and B.~Swiezewska, \emph{{Constraining
  Inert Dark Matter by $R_{\gamma\gamma}$ and WMAP data}},
  \href{http://dx.doi.org/10.1007/JHEP09(2013)055}{\emph{JHEP} {\bf 09} (2013)
  055}, [\href{http://arxiv.org/abs/1305.6266}{{\tt 1305.6266}}].

\bibitem{Osland:2013sla}
P.~Osland, A.~Pukhov, G.~M. Pruna and M.~Purmohammadi, \emph{{Phenomenology of
  charged scalars in the CP-Violating Inert-Doublet Model}},
  \href{http://dx.doi.org/10.1007/JHEP04(2013)040}{\emph{JHEP} {\bf 04} (2013)
  040}, [\href{http://arxiv.org/abs/1302.3713}{{\tt 1302.3713}}].

\bibitem{Abe:2015rja}
T.~Abe and R.~Sato, \emph{{Quantum corrections to the spin-independent cross
  section of the inert doublet dark matter}},
  \href{http://dx.doi.org/10.1007/JHEP03(2015)109}{\emph{JHEP} {\bf 03} (2015)
  109}, [\href{http://arxiv.org/abs/1501.04161}{{\tt 1501.04161}}].

\bibitem{Arhrib:2015hoa}
A.~Arhrib, R.~Benbrik, J.~El~Falaki and A.~Jueid, \emph{{Radiative corrections
  to the Triple Higgs Coupling in the Inert Higgs Doublet Model}},
  \href{http://dx.doi.org/10.1007/JHEP12(2015)007}{\emph{JHEP} {\bf 12} (2015)
  007}, [\href{http://arxiv.org/abs/1507.03630}{{\tt 1507.03630}}].

\bibitem{Blinov:2015qva}
N.~Blinov, J.~Kozaczuk, D.~E. Morrissey and A.~de~la Puente, \emph{{Compressing
  the Inert Doublet Model}},
  \href{http://dx.doi.org/10.1103/PhysRevD.93.035020}{\emph{Phys. Rev.} {\bf
  D93} (2016) 035020}, [\href{http://arxiv.org/abs/1510.08069}{{\tt
  1510.08069}}].

\bibitem{Diaz:2015pyv}
M.~A. Díaz, B.~Koch and S.~Urrutia-Quiroga, \emph{{Constraints to Dark Matter
  from Inert Higgs Doublet Model}},
  \href{http://dx.doi.org/10.1155/2016/8278375}{\emph{Adv. High Energy Phys.}
  {\bf 2016} (2016) 8278375}, [\href{http://arxiv.org/abs/1511.04429}{{\tt
  1511.04429}}].

\bibitem{Ilnicka:2015jba}
A.~Ilnicka, M.~Krawczyk and T.~Robens, \emph{{Inert Doublet Model in light of
  LHC Run I and astrophysical data}},
  \href{http://dx.doi.org/10.1103/PhysRevD.93.055026}{\emph{Phys. Rev.} {\bf
  D93} (2016) 055026}, [\href{http://arxiv.org/abs/1508.01671}{{\tt
  1508.01671}}].

\bibitem{Belanger:2015kga}
G.~Belanger, B.~Dumont, A.~Goudelis, B.~Herrmann, S.~Kraml and D.~Sengupta,
  \emph{{Dilepton constraints in the Inert Doublet Model from Run1 of the
  LHC}}, \href{http://dx.doi.org/10.1103/PhysRevD.91.115011}{\emph{Phys. Rev.}
  {\bf D91} (2015) 115011}, [\href{http://arxiv.org/abs/1503.07367}{{\tt
  1503.07367}}].

\bibitem{Carmona:2015haa}
A.~Carmona and M.~Chala, \emph{{Composite Dark Sectors}},
  \href{http://dx.doi.org/10.1007/JHEP06(2015)105}{\emph{JHEP} {\bf 06} (2015)
  105}, [\href{http://arxiv.org/abs/1504.00332}{{\tt 1504.00332}}].

\bibitem{Kanemura:2016sos}
S.~Kanemura, M.~Kikuchi and K.~Sakurai, \emph{{Testing the dark matter scenario
  in the inert doublet model by future precision measurements of the Higgs
  boson couplings}},
  \href{http://dx.doi.org/10.1103/PhysRevD.94.115011}{\emph{Phys. Rev.} {\bf
  D94} (2016) 115011}, [\href{http://arxiv.org/abs/1605.08520}{{\tt
  1605.08520}}].

\bibitem{Queiroz:2015utg}
F.~S. Queiroz and C.~E. Yaguna, \emph{{The CTA aims at the Inert Doublet
  Model}}, \href{http://dx.doi.org/10.1088/1475-7516/2016/02/038}{\emph{JCAP}
  {\bf 1602} (2016) 038}, [\href{http://arxiv.org/abs/1511.05967}{{\tt
  1511.05967}}].

\bibitem{Belyaev:2016lok}
A.~Belyaev, G.~Cacciapaglia, I.~P. Ivanov, F.~Rojas-Abatte and M.~Thomas,
  \emph{{Anatomy of the Inert Two Higgs Doublet Model in the light of the LHC
  and non-LHC Dark Matter Searches}},
  \href{http://dx.doi.org/10.1103/PhysRevD.97.035011}{\emph{Phys. Rev.} {\bf
  D97} (2018) 035011}, [\href{http://arxiv.org/abs/1612.00511}{{\tt
  1612.00511}}].

\bibitem{Arcadi:2019lka}
G.~Arcadi, A.~Djouadi and M.~Raidal, \emph{{Dark Matter through the Higgs
  portal}}, \href{http://dx.doi.org/10.1016/j.physrep.2019.11.003}{\emph{Phys.
  Rept.} {\bf 842} (2020) 1--180}, [\href{http://arxiv.org/abs/1903.03616}{{\tt
  1903.03616}}].

\bibitem{Eiteneuer:2017hoh}
B.~Eiteneuer, A.~Goudelis and J.~Heisig, \emph{{The inert doublet model in the
  light of Fermi-LAT gamma-ray data: a global fit analysis}},
  \href{http://dx.doi.org/10.1140/epjc/s10052-017-5166-1}{\emph{Eur. Phys. J.}
  {\bf C77} (2017) 624}, [\href{http://arxiv.org/abs/1705.01458}{{\tt
  1705.01458}}].

\bibitem{Ilnicka:2018def}
A.~Ilnicka, T.~Robens and T.~Stefaniak, \emph{{Constraining Extended Scalar
  Sectors at the LHC and beyond}},
  \href{http://dx.doi.org/10.1142/S0217732318300070}{\emph{Mod. Phys. Lett.}
  {\bf A33} (2018) 1830007}, [\href{http://arxiv.org/abs/1803.03594}{{\tt
  1803.03594}}].

\bibitem{Kalinowski:2018ylg}
J.~Kalinowski, W.~Kotlarski, T.~Robens, D.~Sokolowska and A.~F. Zarnecki,
  \emph{{Benchmarking the Inert Doublet Model for $e^+ e^-$ colliders}},
  \href{http://dx.doi.org/10.1007/JHEP12(2018)081}{\emph{JHEP} {\bf 12} (2018)
  081}, [\href{http://arxiv.org/abs/1809.07712}{{\tt 1809.07712}}].

\bibitem{Basu:2020qoe}
R.~Basu, S.~Banerjee, M.~Pandey and D.~Majumdar, \emph{{Lower bounds on dark
  matter annihilation cross-sections by studying the fluctuations of 21-cm line
  with dark matter candidate in inert doublet model (IDM) with the combined
  effects of dark matter scattering and annihilation}},
  \href{http://arxiv.org/abs/2010.11007}{{\tt 2010.11007}}.

\bibitem{Abouabid:2020eik}
H.~Abouabid, A.~Arhrib, R.~Benbrik, J.~E. Falaki, B.~Gong, W.~Xie et~al.,
  \emph{{One-loop radiative corrections to $e^+ e^-\to Zh^0/H^0A^0$ in the
  Inert Higgs Doublet Model}},  \href{http://arxiv.org/abs/2009.03250}{{\tt
  2009.03250}}.

\bibitem{Ferreira:2009jb}
P.~M. Ferreira and D.~R.~T. Jones, \emph{{Bounds on scalar masses in two Higgs
  doublet models}},
  \href{http://dx.doi.org/10.1088/1126-6708/2009/08/069}{\emph{JHEP} {\bf 08}
  (2009) 069}, [\href{http://arxiv.org/abs/0903.2856}{{\tt 0903.2856}}].

\bibitem{Ferreira:2015pfi}
P.~M. Ferreira and B.~Swiezewska, \emph{{One-loop contributions to neutral
  minima in the inert doublet model}},
  \href{http://dx.doi.org/10.1007/JHEP04(2016)099}{\emph{JHEP} {\bf 04} (2016)
  099}, [\href{http://arxiv.org/abs/1511.02879}{{\tt 1511.02879}}].

\bibitem{Kanemura:2002vm}
S.~Kanemura, S.~Kiyoura, Y.~Okada, E.~Senaha and C.~P. Yuan, \emph{{New physics
  effect on the Higgs selfcoupling}},
  \href{http://dx.doi.org/10.1016/S0370-2693(03)00268-5}{\emph{Phys. Lett.}
  {\bf B558} (2003) 157--164}, [\href{http://arxiv.org/abs/hep-ph/0211308}{{\tt
  hep-ph/0211308}}].

\bibitem{Senaha:2018xek}
E.~Senaha, \emph{{Radiative Corrections to Triple Higgs Coupling and
  Electroweak Phase Transition: Beyond One-loop Analysis}},
  \href{http://dx.doi.org/10.1103/PhysRevD.100.055034}{\emph{Phys. Rev. D} {\bf
  100} (2019) 055034}, [\href{http://arxiv.org/abs/1811.00336}{{\tt
  1811.00336}}].

\bibitem{Braathen:2019pxr}
J.~Braathen and S.~Kanemura, \emph{{On two-loop corrections to the Higgs
  trilinear coupling in models with extended scalar sectors}},
  \href{http://dx.doi.org/10.1016/j.physletb.2019.07.021}{\emph{Phys. Lett. B}
  {\bf 796} (2019) 38--46}, [\href{http://arxiv.org/abs/1903.05417}{{\tt
  1903.05417}}].

\bibitem{Arhrib:2012ia}
A.~Arhrib, R.~Benbrik and N.~Gaur, \emph{{$H\to \gamma \gamma$ in Inert Higgs
  Doublet Model}},
  \href{http://dx.doi.org/10.1103/PhysRevD.85.095021}{\emph{Phys. Rev.} {\bf
  D85} (2012) 095021}, [\href{http://arxiv.org/abs/1201.2644}{{\tt
  1201.2644}}].

\bibitem{Gustafsson:2007pc}
M.~Gustafsson, E.~Lundstrom, L.~Bergstrom and J.~Edsjo, \emph{{Significant
  Gamma Lines from Inert Higgs Dark Matter}},
  \href{http://dx.doi.org/10.1103/PhysRevLett.99.041301}{\emph{Phys. Rev.
  Lett.} {\bf 99} (2007) 041301},
  [\href{http://arxiv.org/abs/astro-ph/0703512}{{\tt astro-ph/0703512}}].

\bibitem{Garcia-Cely:2015khw}
C.~Garcia-Cely, M.~Gustafsson and A.~Ibarra, \emph{{Probing the Inert Doublet
  Dark Matter Model with Cherenkov Telescopes}},
  \href{http://dx.doi.org/10.1088/1475-7516/2016/02/043}{\emph{JCAP} {\bf 1602}
  (2016) 043}, [\href{http://arxiv.org/abs/1512.02801}{{\tt 1512.02801}}].

\bibitem{Banerjee:2016vrp}
S.~Banerjee and N.~Chakrabarty, \emph{{A revisit to scalar dark matter with
  radiative corrections}},
  \href{http://dx.doi.org/10.1007/JHEP05(2019)150}{\emph{JHEP} {\bf 05} (2019)
  150}, [\href{http://arxiv.org/abs/1612.01973}{{\tt 1612.01973}}].

\bibitem{Boudjema:1995cb}
F.~Boudjema and E.~Chopin, \emph{{Double Higgs production at the linear
  colliders and the probing of the Higgs selfcoupling}},
  \href{http://dx.doi.org/10.1007/s002880050298}{\emph{Z. Phys. C} {\bf 73}
  (1996) 85--110}, [\href{http://arxiv.org/abs/hep-ph/9507396}{{\tt
  hep-ph/9507396}}].

\bibitem{Baro:2007em}
N.~Baro, F.~Boudjema and A.~Semenov, \emph{{Full one-loop corrections to the
  relic density in the MSSM: A Few examples}},
  \href{http://dx.doi.org/10.1016/j.physletb.2008.01.031}{\emph{Phys. Lett.}
  {\bf B660} (2008) 550--560}, [\href{http://arxiv.org/abs/0710.1821}{{\tt
  0710.1821}}].

\bibitem{OurPaper4_2020}
S.~Banerjee, F.~Boudjema, N.~Chakrabarty and H.~Sun, \emph{{Relic density of
  dark matter in the inert doublet model beyond leading order: The SM Higgs
  resonance region}}, .

\bibitem{OurPaper2_2020}
S.~Banerjee, F.~Boudjema, N.~Chakrabarty and H.~Sun, \emph{{Relic density of
  dark matter in the inert doublet model beyond leading order: Co-annihilation
  in the low mass region}}, .

\bibitem{OurPaper3_2020}
S.~Banerjee, F.~Boudjema, N.~Chakrabarty and H.~Sun, \emph{{Relic density of
  dark matter in the inert doublet model beyond leading order: Annihilation in
  3-body final state for the low mass region}}, .

\bibitem{Boudjema_2005}
F.~Boudjema, A.~Semenov and D.~Temes, \emph{Self-annihilation of the neutralino
  dark matter into two photons or azand a photon in the minimal supersymmetric
  standard model},
  \href{http://dx.doi.org/10.1103/physrevd.72.055024}{\emph{Physical Review D}
  {\bf 72} (Sep, 2005) }.

\bibitem{Baro:2008bg}
N.~Baro, F.~Boudjema and A.~Semenov, \emph{{Automatised full one-loop
  renormalisation of the MSSM. I. The Higgs sector, the issue of tan(beta) and
  gauge invariance}},
  \href{http://dx.doi.org/10.1103/PhysRevD.78.115003}{\emph{Phys. Rev.} {\bf
  D78} (2008) 115003}, [\href{http://arxiv.org/abs/0807.4668}{{\tt
  0807.4668}}].

\bibitem{Baro:2009na}
N.~Baro, F.~Boudjema, G.~Chalons and S.~Hao, \emph{{Relic density at one-loop
  with gauge boson pair production}},
  \href{http://dx.doi.org/10.1103/PhysRevD.81.015005}{\emph{Phys. Rev.} {\bf
  D81} (2010) 015005}, [\href{http://arxiv.org/abs/0910.3293}{{\tt
  0910.3293}}].

\bibitem{Boudjema:2011ig}
F.~Boudjema, G.~Drieu La~Rochelle and S.~Kulkarni, \emph{{One-loop corrections,
  uncertainties and approximations in neutralino annihilations: Examples}},
  \href{http://dx.doi.org/10.1103/PhysRevD.84.116001}{\emph{Phys. Rev. D} {\bf
  84} (2011) 116001}, [\href{http://arxiv.org/abs/1108.4291}{{\tt 1108.4291}}].

\bibitem{Boudjema:2014gza}
F.~Boudjema, G.~Drieu La~Rochelle and A.~Mariano, \emph{{Relic density
  calculations beyond tree-level, exact calculations versus effective
  couplings: the ZZ final state}},
  \href{http://dx.doi.org/10.1103/PhysRevD.89.115020}{\emph{Phys. Rev.} {\bf
  D89} (2014) 115020}, [\href{http://arxiv.org/abs/1403.7459}{{\tt
  1403.7459}}].

\bibitem{Belanger:2016tqb}
G.~B\'elanger, V.~Bizouard, F.~Boudjema and G.~Chalons, \emph{{One-loop
  renormalization of the NMSSM in SloopS: The neutralino-chargino and sfermion
  sectors}}, \href{http://dx.doi.org/10.1103/PhysRevD.93.115031}{\emph{Phys.
  Rev. D} {\bf 93} (2016) 115031}, [\href{http://arxiv.org/abs/1602.05495}{{\tt
  1602.05495}}].

\bibitem{Belanger:2017rgu}
G.~B\'elanger, V.~Bizouard, F.~Boudjema and G.~Chalons, \emph{{One-loop
  renormalization of the NMSSM in SloopS. II. The Higgs sector}},
  \href{http://dx.doi.org/10.1103/PhysRevD.96.015040}{\emph{Phys. Rev.} {\bf
  D96} (2017) 015040}, [\href{http://arxiv.org/abs/1705.02209}{{\tt
  1705.02209}}].

\bibitem{Hahn:2000kx}
T.~Hahn, \emph{{Generating Feynman diagrams and amplitudes with FeynArts 3}},
  \href{http://dx.doi.org/10.1016/S0010-4655(01)00290-9}{\emph{Comput. Phys.
  Commun.} {\bf 140} (2001) 418--431},
  [\href{http://arxiv.org/abs/hep-ph/0012260}{{\tt hep-ph/0012260}}].

\bibitem{Hahn_2016}
T.~Hahn, S.~Passehr and C.~Schappacher, \emph{Formcalc 9 and extensions},
  \href{http://dx.doi.org/10.1088/1742-6596/762/1/012065}{\emph{Journal of
  Physics: Conference Series} {\bf 762} (Oct, 2016) 012065}.

\bibitem{Hahn:1998yk}
T.~Hahn and M.~Perez-Victoria, \emph{{Automatized one loop calculations in
  four-dimensions and D-dimensions}},
  \href{http://dx.doi.org/10.1016/S0010-4655(98)00173-8}{\emph{Comput. Phys.
  Commun.} {\bf 118} (1999) 153--165},
  [\href{http://arxiv.org/abs/hep-ph/9807565}{{\tt hep-ph/9807565}}].

\bibitem{Semenov:2008jy}
A.~Semenov, \emph{{LanHEP: A Package for the automatic generation of Feynman
  rules in field theory. Version 3.0}},
  \href{http://dx.doi.org/10.1016/j.cpc.2008.10.012}{\emph{Comput. Phys.
  Commun.} {\bf 180} (2009) 431--454},
  [\href{http://arxiv.org/abs/0805.0555}{{\tt 0805.0555}}].

\bibitem{Semenov:2014rea}
A.~Semenov, \emph{{LanHEP — A package for automatic generation of Feynman
  rules from the Lagrangian. Version 3.2}},
  \href{http://dx.doi.org/10.1016/j.cpc.2016.01.003}{\emph{Comput. Phys.
  Commun.} {\bf 201} (2016) 167--170},
  [\href{http://arxiv.org/abs/1412.5016}{{\tt 1412.5016}}].

\bibitem{Belanger:2001fz}
G.~Belanger, F.~Boudjema, A.~Pukhov and A.~Semenov, \emph{{MicrOMEGAs: A
  Program for calculating the relic density in the MSSM}},
  \href{http://dx.doi.org/10.1016/S0010-4655(02)00596-9}{\emph{Comput. Phys.
  Commun.} {\bf 149} (2002) 103--120},
  [\href{http://arxiv.org/abs/hep-ph/0112278}{{\tt hep-ph/0112278}}].

\bibitem{Belanger:2006is}
G.~Belanger, F.~Boudjema, A.~Pukhov and A.~Semenov, \emph{{MicrOMEGAs 2.0: A
  Program to calculate the relic density of dark matter in a generic model}},
  \href{http://dx.doi.org/10.1016/j.cpc.2006.11.008}{\emph{Comput. Phys.
  Commun.} {\bf 176} (2007) 367--382},
  [\href{http://arxiv.org/abs/hep-ph/0607059}{{\tt hep-ph/0607059}}].

\bibitem{Belanger:2013oya}
G.~Belanger, F.~Boudjema, A.~Pukhov and A.~Semenov, \emph{{micrOMEGAs$\_$3: A
  program for calculating dark matter observables}},
  \href{http://dx.doi.org/10.1016/j.cpc.2013.10.016}{\emph{Comput. Phys.
  Commun.} {\bf 185} (2014) 960--985},
  [\href{http://arxiv.org/abs/1305.0237}{{\tt 1305.0237}}].

\bibitem{Belanger:2018mqt}
G.~Belanger, F.~Boudjema, A.~Goudelis, A.~Pukhov and B.~Zaldivar,
  \emph{{micrOMEGAs5.0 : Freeze-in}},
  \href{http://dx.doi.org/10.1016/j.cpc.2018.04.027}{\emph{Comput. Phys.
  Commun.} {\bf 231} (2018) 173--186},
  [\href{http://arxiv.org/abs/1801.03509}{{\tt 1801.03509}}].

\bibitem{Haber:1993an}
H.~E. Haber and R.~Hempfling, \emph{{The Renormalization group improved Higgs
  sector of the minimal supersymmetric model}},
  \href{http://dx.doi.org/10.1103/PhysRevD.48.4280}{\emph{Phys. Rev. D} {\bf
  48} (1993) 4280--4309}, [\href{http://arxiv.org/abs/hep-ph/9307201}{{\tt
  hep-ph/9307201}}].

\bibitem{Khan:2015ipa}
N.~Khan and S.~Rakshit, \emph{{Constraints on inert dark matter from the
  metastability of the electroweak vacuum}},
  \href{http://dx.doi.org/10.1103/PhysRevD.92.055006}{\emph{Phys. Rev. D} {\bf
  92} (2015) 055006}, [\href{http://arxiv.org/abs/1503.03085}{{\tt
  1503.03085}}].

\bibitem{Kalinowski:2020rmb}
J.~Kalinowski, T.~Robens, D.~Sokolowska and A.~F. Zarnecki, \emph{{IDM
  benchmarks for the LHC and future colliders}},
  \href{http://arxiv.org/abs/2012.14818}{{\tt 2012.14818}}.

\bibitem{Aprile:2018dbl}
{\scshape XENON} collaboration, E.~Aprile et~al., \emph{{Dark Matter Search
  Results from a One Ton-Year Exposure of XENON1T}},
  \href{http://dx.doi.org/10.1103/PhysRevLett.121.111302}{\emph{Phys. Rev.
  Lett.} {\bf 121} (2018) 111302}, [\href{http://arxiv.org/abs/1805.12562}{{\tt
  1805.12562}}].

\bibitem{Aaboud:2018puo}
{\scshape ATLAS} collaboration, M.~Aaboud et~al., \emph{{Constraints on
  off-shell Higgs boson production and the Higgs boson total width in
  $ZZ\to4\ell$ and $ZZ\to2\ell2\nu$ final states with the ATLAS detector}},
  \href{http://dx.doi.org/10.1016/j.physletb.2018.09.048}{\emph{Phys. Lett. B}
  {\bf 786} (2018) 223--244}, [\href{http://arxiv.org/abs/1808.01191}{{\tt
  1808.01191}}].

\bibitem{Sirunyan:2019twz}
{\scshape CMS} collaboration, A.~M. Sirunyan et~al., \emph{{Measurements of the
  Higgs boson width and anomalous $HVV$ couplings from on-shell and off-shell
  production in the four-lepton final state}},
  \href{http://dx.doi.org/10.1103/PhysRevD.99.112003}{\emph{Phys. Rev. D} {\bf
  99} (2019) 112003}, [\href{http://arxiv.org/abs/1901.00174}{{\tt
  1901.00174}}].

\bibitem{Kraml:2019sis}
S.~Kraml, T.~Q. Loc, D.~T. Nhung and L.~D. Ninh, \emph{{Constraining new
  physics from Higgs measurements with Lilith: update to LHC Run 2 results}},
  \href{http://dx.doi.org/10.21468/SciPostPhys.7.4.052}{\emph{SciPost Phys.}
  {\bf 7} (2019) 052}, [\href{http://arxiv.org/abs/1908.03952}{{\tt
  1908.03952}}].

\bibitem{10.1093/ptep/ptaa104}
P.~D. Group, \emph{{Review of Particle Physics}},
  \href{http://dx.doi.org/10.1093/ptep/ptaa104}{\emph{Progress of Theoretical
  and Experimental Physics} {\bf 2020} (08, 2020) },
  [\href{http://arxiv.org/abs/See in particular, pages 13 and 14 online
  http://pdg.lbl.gov/2019/reviews/rpp2019-rev-higgs-boson.pdf}{{\tt See in
  particular, pages 13 and 14 online
  http://pdg.lbl.gov/2019/reviews/rpp2019-rev-higgs-boson.pdf}}].

\bibitem{deFlorian:2016spz}
{\scshape LHC Higgs Cross Section Working Group} collaboration, D.~de~Florian
  et~al., \emph{{Handbook of LHC Higgs Cross Sections: 4. Deciphering the
  Nature of the Higgs Sector. arXiv: 1610.07922}},
  \href{http://arxiv.org/abs/1610.07922}{{\tt 1610.07922}}.

\bibitem{Peskin:1991sw}
M.~E. Peskin and T.~Takeuchi, \emph{{Estimation of oblique electroweak
  corrections}}, \href{http://dx.doi.org/10.1103/PhysRevD.46.381}{\emph{Phys.
  Rev. D} {\bf 46} (1992) 381--409}.

\bibitem{Baak_2012}
M.~Baak, M.~Goebel, J.~Haller, A.~Hoecker, D.~Kennedy, K.~Mönig et~al.,
  \emph{Updated status of the global electroweak fit and constraints on new
  physics}, \href{http://dx.doi.org/10.1140/epjc/s10052-012-2003-4}{\emph{The
  European Physical Journal C} {\bf 72} (May, 2012) }.

\bibitem{Pierce:2007ut}
A.~Pierce and J.~Thaler, \emph{{Natural Dark Matter from an Unnatural Higgs
  Boson and New Colored Particles at the TeV Scale}},
  \href{http://dx.doi.org/10.1088/1126-6708/2007/08/026}{\emph{JHEP} {\bf 08}
  (2007) 026}, [\href{http://arxiv.org/abs/hep-ph/0703056}{{\tt
  hep-ph/0703056}}].

\bibitem{Datta:2016nfz}
A.~Datta, N.~Ganguly, N.~Khan and S.~Rakshit, \emph{{Exploring collider
  signatures of the inert Higgs doublet model}},
  \href{http://dx.doi.org/10.1103/PhysRevD.95.015017}{\emph{Phys. Rev.} {\bf
  D95} (2017) 015017}, [\href{http://arxiv.org/abs/1610.00648}{{\tt
  1610.00648}}].

\bibitem{Dolle_2010}
E.~Dolle, X.~Miao, S.~Su and B.~Thomas, \emph{Dilepton signals in the inert
  doublet model},
  \href{http://dx.doi.org/10.1103/physrevd.81.035003}{\emph{Physical Review D}
  {\bf 81} (Feb, 2010) }.

\bibitem{LHCXSWG}
\emph{{\uppercase{LHC SUSY}} cross section working group. \\
  https://twiki.cern.ch/twiki/bin/view/lhcphysics/susycrosssections}, .

\bibitem{Dumont:2014tja}
B.~Dumont, B.~Fuks, S.~Kraml, S.~Bein, G.~Chalons, E.~Conte et~al.,
  \emph{{Toward a public analysis database for LHC new physics searches using
  MADANALYSIS 5}},
  \href{http://dx.doi.org/10.1140/epjc/s10052-014-3242-3}{\emph{Eur. Phys. J.
  C} {\bf 75} (2015) 56}, [\href{http://arxiv.org/abs/1407.3278}{{\tt
  1407.3278}}].

\bibitem{Conte:2018vmg}
E.~Conte and B.~Fuks, \emph{{Confronting new physics theories to LHC data with
  MADANALYSIS 5}},
  \href{http://dx.doi.org/10.1142/S0217751X18300272}{\emph{Int. J. Mod. Phys.
  A} {\bf 33} (2018) 1830027}, [\href{http://arxiv.org/abs/1808.00480}{{\tt
  1808.00480}}].

\bibitem{Araz:2020lnp}
J.~Y. Araz, B.~Fuks and G.~Polykratis, \emph{{Simplified fast detector
  simulation in MadAnalysis 5}},  \href{http://arxiv.org/abs/2006.09387}{{\tt
  2006.09387}}.

\bibitem{Aad:2019vnb}
{\scshape ATLAS} collaboration, G.~Aad et~al., \emph{{Search for electroweak
  production of charginos and sleptons decaying into final states with two
  leptons and missing transverse momentum in $\sqrt{s}=13$ TeV $pp$ collisions
  using the ATLAS detector}},
  \href{http://dx.doi.org/10.1140/epjc/s10052-019-7594-6}{\emph{Eur. Phys. J.
  C} {\bf 80} (2020) 123}, [\href{http://arxiv.org/abs/1908.08215}{{\tt
  1908.08215}}].

\bibitem{DVN/EA4S4D_2020}
B.~Fuks and J.~Y. Araz, \emph{{Re-implementation of a search for sleptons and
  electroweakinos in the dilepton + MET channel (139 fb $^{-1}$;
  ATLAS-SUSY-2018-32)}}, .

\bibitem{Sirunyan:2017lae}
{\scshape CMS} collaboration, A.~Sirunyan et~al., \emph{{Search for electroweak
  production of charginos and neutralinos in multilepton final states in
  proton-proton collisions at $\sqrt{s}=$ 13 TeV}},
  \href{http://dx.doi.org/10.1007/JHEP03(2018)166}{\emph{JHEP} {\bf 03} (2018)
  166}, [\href{http://arxiv.org/abs/1709.05406}{{\tt 1709.05406}}].

\bibitem{electroweakinos}
B.~Fuks and S.~Mondal, \emph{{MadAnalysis 5 implementation of the CMS search
  for supersymmetry in the multilepton channel with 35.9~fb~$^{-1}$ of 13 TeV
  LHC data (CMS-SUS-16-039)}}, .

\bibitem{Sirunyan:2017onm}
{\scshape CMS} collaboration, A.~M. Sirunyan et~al., \emph{{Search for dark
  matter and unparticles in events with a Z boson and missing transverse
  momentum in proton-proton collisions at $ \sqrt{s}=13 $ TeV}},
  \href{http://dx.doi.org/10.1007/JHEP03(2017)061}{\emph{JHEP} {\bf 03} (2017)
  061}, [\href{http://arxiv.org/abs/1701.02042}{{\tt 1701.02042}}].

\bibitem{monoZ}
B.~Fuks, \emph{{MadAnalysis5 implementation of the mono-Z analysis of CMS with
  2.3 fb$^{-1}$ of data (CMS-EXO-16-010)}}, .

\bibitem{Aad:2019wvl}
{\scshape ATLAS} collaboration, G.~Aad et~al., \emph{{Search for a heavy
  charged boson in events with a charged lepton and missing transverse momentum
  from $pp$ collisions at $\sqrt{s} = 13$ TeV with the ATLAS detector}},
  \href{http://dx.doi.org/10.1103/PhysRevD.100.052013}{\emph{Phys. Rev. D} {\bf
  100} (2019) 052013}, [\href{http://arxiv.org/abs/1906.05609}{{\tt
  1906.05609}}].

\bibitem{DVN/GLWLTF_2020}
K.~Park, S.~Lee, W.~Jun and U.~Min, \emph{{Re-implementation of the W' into 1
  lepton + missing energy analysis (139 fb-1; ATLAS-EXOT-2018-30)}}, .

\bibitem{Araz:2019otb}
J.~Y. Araz, M.~Frank and B.~Fuks, \emph{{Reinterpreting the results of the LHC
  with MadAnalysis 5: uncertainties and higher-luminosity estimates}},
  \href{http://dx.doi.org/10.1140/epjc/s10052-020-8076-6}{\emph{Eur. Phys. J.
  C} {\bf 80} (2020) 531}, [\href{http://arxiv.org/abs/1910.11418}{{\tt
  1910.11418}}].

\bibitem{PhysRevD.98.052005}
{\scshape ATLAS Collaboration} collaboration, M.~{\it et al.}. Aaboud,
  \emph{Measurements of higgs boson properties in the diphoton decay channel
  with $36\text{ }\text{ }{\mathrm{fb}}^{\ensuremath{-}1}$ of $pp$ collision
  data at $\sqrt{s}=13\text{ }\text{ }\mathrm{TeV}$ with the atlas detector},
  \href{http://dx.doi.org/10.1103/PhysRevD.98.052005}{\emph{Phys. Rev. D} {\bf
  98} (Sep, 2018) 052005}.

\bibitem{Sirunyan_2018}
{\scshape CMS Collaboration} collaboration, A.~{\it et al.,}. Sirunyan,
  \emph{Measurements of higgs boson properties in the diphoton decay channel in
  proton-proton collisions at $\sqrt{s}=13\text{ }\text{ }\mathrm{TeV}$},
  \href{http://dx.doi.org/10.1007/jhep11(2018)185}{\emph{Journal of High Energy
  Physics} {\bf 2018} (Nov, 2018) }.

\bibitem{Alwall_2014}
J.~Alwall, R.~Frederix, S.~Frixione, V.~Hirschi, F.~Maltoni, O.~Mattelaer
  et~al., \emph{The automated computation of tree-level and next-to-leading
  order differential cross sections, and their matching to parton shower
  simulations}, \href{http://dx.doi.org/10.1007/jhep07(2014)079}{\emph{Journal
  of High Energy Physics} {\bf 2014} (Jul, 2014) }.

\bibitem{Belanger:2004yn}
G.~Belanger, F.~Boudjema, A.~Pukhov and A.~Semenov, \emph{{micrOMEGAs: Version
  1.3}}, \href{http://dx.doi.org/10.1016/j.cpc.2005.12.005}{\emph{Comput. Phys.
  Commun.} {\bf 174} (2006) 577--604},
  [\href{http://arxiv.org/abs/hep-ph/0405253}{{\tt hep-ph/0405253}}].

\end{thebibliography}\endgroup
\bibliographystyle{JHEP}

\end{document}